\documentclass[12pt]{iopart}
\usepackage{graphicx,subfig,url}
\usepackage{iopams}
\begin{document}
\title[Validation of the thermo-chemical model of CO$_2$ plasma conversion]
{Validation of the thermo-chemical approach to modelling of the CO$_2$ 
conversion in atmospheric pressure microwave gas discharges}
\author{Vladislav Kotov$^1$, Christian Kiefer$^2$ and Ante Hecimovic$^2$}
\address{$^1$ Forschungszentrum J\"ulich GmbH, Institut f\"ur Energie- und Klimaforschung - Plasmaphysik (IEK-4), 
              Partner of the Trilateral Euregio Cluster (TEC), 52425, J\"ulich, Germany}
\address{$^2$ Max-Planck-Institut f\"ur  Plasmaphysik, Plasma for Gas conversion (P4G) group - ITED, 85748, Garching, Germany}

\ead{v.kotov@fz-juelich.de}

\date{\today}

\begin{abstract} 
The 2.45~GHz plasma torch CO$_2\to$~CO~+~$\frac12$O$_2$ conversion experiment 
[F~A~D’Isa et al. 2020 {\it Plasma Sources Sci. Technol.}{\bf 29} 105009] 
has been compared with thermo-chemical calculations. 
The 1.5D model of the CO$_2$/CO/O$_2$/O/C mixture without turbulent transport has been used with plasma acting only as prescribed heat source. 
The parameter range covered is specific energy input SEI=0.3..5~eV/molecule at pressure $p$=0.9~bar, and 
SEI=0.6..2~eV/molecule at $p$=0.5, 0.2~bar.       
The calculated conversion rates $\chi$ are always in good agreement with experiment.
At the same time, the calculated temperatures $T$ may deviate significantly from the experiment, especially for $p$=0.2~bar.            
The calculated $T$ were found to be sensitive with respect to uncertain model parameters, but $\chi$ are not sensitive. 
The model suggests that the conversion process is essentially two-dimensional and that the main factor which reduces the 
energy efficiency is re-oxidation of CO downstream from the plasma region. 
The proposed physico-chemical model (chemical mechanism plus transport coefficients) can be suggested 
for practical calculations for $p\geq$0.5~bar.
\end{abstract}  

\noindent{\it microwave discharges, plasma torch, computer modeling, carbon dioxide, chemical kinetics \/}

\maketitle

\section{Introduction}

Plasma chemical conversion of CO$_2$ into CO has a potential to become an important part of the future 
Carbon Capture and Utilization process chains~\cite{BogaertsNeyts2018}.
Remarkable progress has been achieved in this field in the recent years in particular in 
microwave (MW) flowing plasma reactors where the chemical energy efficiencies
(efficiencies related to the energy absorbed in plasma) 
of the conversions process up to 40..50~\% were achieved~\cite{Bongers2017,denHarder2017,DIsa2020,Hecimovic2022}. 
However, further increase of efficiency and productivity -  especially when operating at atmospheric pressure -  
is required to make this technology mature for practical applications. 

It has been demonstrated experimentally that in this type of reactors 
modifying the flow in the discharge tube by installing nozzles and orifices 
as well as by using different gas supply schemes 
have significant impact on the process performance~\cite{Bongers2017,Hecimovic2022}.
Thus, it may be possible to further optimize the conversion process by influencing the flow characteristics. 
A purely experimental optimization would require building a large number of prototypes and executing a large 
amount of experiments which may be prohibitive in terms of time and resources consumption. 
A common practice for reducing the number of prototypes and accelerating the process optimization 
is extensive use of numerical modelling for planning and evaluation of experiments. 

A full numerical description of the CO$_2$ conversion in a MW plasma reactor must couple self consistently 
gas flow with chemical reactions, plasma processes and electromagnetic field. 
At least in 2D, ideally in 3D. 
Apparently, such a full self-consistent modelling poses serious difficulties even to modern simulation tools running on modern computers.
For MW discharges 3D models have been so far reported only for Ar plasma~\cite{Baeva2018,Shen2022}. 
2D modelling studies in plasmas sustained by radio-frequency waves including chemical reactions 
have been done for N$_2$~\cite{MingaoYu2016,MingaoYu2021}.
Recently, a fully self-consistent 2D modelling of the CO$_2$ plasma chemical conversion in MW reactor 
have been published~\cite{Elahi2023}, however, only for CO$_2$ diluted in Ar in the proportion 1:7. 
Most relevant for practical applications is the conversion at atmospheric pressure which does not require vacuum equipment. 
Fortunately, it turns out that in that case practically relevant results can be achieved already by much more simple 
non-self-consistent models. 

Particularly extensive experimental studies of the CO$_2$ conversion in MW plasmas
in pressures ranging from 1~mbar to 1~bar have been 
performed in the devices with 2.45~GHz TE$_{10}$ waves - surfaguides, plasma torches and similar configurations. 
Analysis of the plasma-chemical mechanisms in weekly ionized plasma of that type of reactors 
led to conclusion that in the medium to atmospheric pressure range ($\geq$200~mbar) 
reactions with charged particles play minor role in the net chemical process~\cite{Viegas2020,vandeSteeg2021,Vialetto2022}. 
The observed conversion rates can be largely explained by chemical non-equilibrium 
where the plasma only plays a role of the heat source~\cite{Bongers2017,denHarder2017,vandenBekerom2019,Wolf2020a}. 
Moreover, in the pressure range in question the plasma is found to be contracted near the axis 
and the localization of this heat source in experiment can be reconstructed: 
by applying the impedance matching technique~\cite{Groen2019}, 
and from the measured radiation intensity~\cite{denHarder2017,DIsa2020,Wolf2019,Wolf2020b,Viegas2021,DIsaThesis2021}. 

All the observations summarized above open the possibility to use purely thermo-chemical approach  
as a first stage of optimization of the nozzle and gas management configurations. 
The term 'thermo-chemical' is applied here to the models where only the gas flow with chemical reactions 
is simulated. The plasma processes are not modelled explicitly and the plasma only enters a as 
prescribed heat source. In such models equations for the mass, energy and momentum (Navier-Stokes equation) conservation of the bulk flow 
are solved together with drift-diffusion equations with finite rate chemistry for each chemical component. 
The thermo-chemical models are well known and widely used in chemical engineering 
and in studies of combustion processes~\cite{ReactingFlowBook2003}.
The task of describing the chemical reactions is further simplified by the fact that 
in MW plasmas with relevant parameters no significant non-equilibrium between vibrational and translational-rotational 
temperatures have been detected~\cite{vandeSteeg2021,Carbone2020}. 
Thus, one can rely on the readily available data on the 
reaction rate coefficients determined for conditions of temperature equilibrium. 

The main goal of the present paper is to validate the thermo-chemical approach by benchmarking it against the 
plasma torch experiment of F.~A.~D'Isa~et~al.~\cite{DIsa2020,DIsaThesis2021}. 
One method of the thermo-chemical analysis which can be found in the literature is 
modelling of the 2D or 3D single fluid flow without chemical reactions, and then applying 1D chemical kinetics models 
along flux tubes~\cite{Yang2018,vanAlphen2023}. In this case only the calculated temperature can be compared with experiment, 
and the local 1D analysis gives only qualitative insight onto the interplay of the bulk flow and the chemical process.
In the present paper another concept is used which goes back to A.~J.~Wolf~et~al.~\cite{Wolf2020a}. 
It is based on applying a simplified model for the bulk flow in the cylindrical discharge tube, 
but solved self-consistently with the set of transport equations for the individual components 
of the CO$_2$/CO/O$_2$/O/C mixture including chemical reactions.  
This approach may be less accurate in capturing the bulk flow pattern and the temperature field, 
but it allows to calculate the target quantities - the conversion rate and energy efficiency - 
and compare them directly with experiment. 

In the present work the emphasis in the validation is made on reproducing trends: 
dependencies of the target quantities on power and flow rate as well as on pressure, rather than trying to 
match as good as possible each individual experiment. Therefore, a simplified 1.5D model is applied. 
Equations for heat balance and drift-diffusion equations are solved in 2D assuming axial symmetry. 
To calculate the bulk flow velocity a local 1D approximation is used instead of the full Navier-Stokes equation. 
This approximation is based on the assumptions of constant static pressure and zero radial mass transfer. 
Whereas the first assumption is well justified by Mach numbers $\ll$1, the second one, strictly speaking, 
has no solid justification. 
This latter assumption implies that the effect of convective cells - if they form - on the net chemical process 
is taken to be negligible. Certain arguments in favor of this hypothesis can be found in~\cite{Wolf2020a} where 
this same assumption was used as well. Another flow feature explicitly omitted here is the tangential swirl 
which is always applied in the discharges in question for stabilization and could potentially 
affect the particles diffusion via centrifugal effect. 
Discussion of those both subjects is left beyond the scope of the present paper. 

Only molecular diffusion and heat conduction coefficients are applied without considering turbulent transport.
This is the main difference of the method of the present work compared to~\cite{Wolf2020a} 
where the turbulent viscosity was used as adjustable parameter to match the measured temperature. 
Here it was not tried to enforce the solution to match the experimental temperature. 
Instead, since the procedure of determining the localization of the plasma zone in experiment is not accurate 
the sensitivity with respect to assumed distribution of the heat source is investigated. 
The rationale for neglecting the turbulent transport as the first approximation 
is that at high gas temperatures considered (up to 6000~K) 
the purely molecular transport is strong and the dominance of the turbulent transport in this case is 
not apparent. Indeed, it will be shown that the applied laminar model can reproduce the measured 
conversion rates very well.  At the same time, the calculated gas temperature may deviate strongly from 
the experimental values. This latter discrepancy could be attributed to the absence of turbulent transport in the model, 
although recent work on CFD with turbulence equations exhibited similar discrepancy in the temperature profile
at 0.2~bar~\cite{Mashayekh2023}. 
However, the temperature was also found to be very sensitive with respect to the assumptions on the spatial distribution of heat source.
Opposite to that, as it will be shown the calculated conversion rates are not sensitive with respect to those assumptions. 

The rest of the paper is organized as follows. In the next section~\ref{description:model:S} the model is described: 
the set of equations solved, the chemical mechanism, and the transport coefficients. 
In section~\ref{comparison:with:experiment:S}, comparison of the model with the experiment~\cite{DIsa2020,DIsaThesis2021} is discussed.
Section~\ref{sensitivity:S} presents the assessment of sensitivity with respect to the  uncertain model parameters, 
and section~\ref{insights:S} gives some insights into 
the atmospheric pressure conversion process provided by the model. Last section~\ref{conclusions:S} summarizes the main conclusions.

\section{Description of the model}

\label{description:model:S}

It is convenient to divide the model into two blocks: first is the transport equations and 
second is the calculation of the coefficients of those equations - the 'physico-chemical model'. 
This division reflects the software-technical implementation. 

\subsection{Transport equations}

The set of transport equations applied here consists of the continuity equations for each chemical component, 
energy balance equation and a simplified local 1D model for axial velocity. 
Only steady state solutions are considered. 

The continuity equations - convection-diffusion equations - read:
\begin{equation}
\label{diffusion:Eq}
{\rm div}{\left( \vec{ \Gamma_i } \right) } = S_i,\qquad
\vec{\Gamma_i} = n\vec{v}X_i - nD_{im}\nabla{X_i},\qquad X_i = \frac{n_i}{n}
\end{equation}
Where $\vec{v}$ is the average mass flow velocity, $n_i$ is the number density of the component $i$, 
$S_i$ is the specific volumetric particle source of the component $i$: $n=\sum_i n_i$. 
Here end below $\sum_i$ always stands for the sum taken over all chemical components. 
In~\eref{diffusion:Eq} the diffusive flux is approximated by Fick's law with the effective mixture-averaged diffusion 
coefficient $D_{im}$, rather than calculated by solving the 
full set of Stefan-Maxwell equations (see~\cite{ReactingFlowBook2003}, 
Chapter 3.5 there and~\cite{CurtissHirschfelder1949}). 
The thermal diffusion and the pressure gradient, 
(see~\cite{ReactingFlowBook2003}, $\S$3.5.2), are neglected. 
This latter would contain the centrifugal effect of the swirl flow on the diffusion. 

The following equation is solved for the gas temperature $T$ - 
which is the common temperature of translational, rotational and vibrational modes which are assumed to be in equilibrium:
\begin{equation}
\label{heat:Eq}
{\rm div}{\left( -\lambda\nabla{T} + \sum_i h_i \vec{\Gamma_i} \right) } = Q - S_h 
\end{equation}
This equation is a reduced form of the total energy conservation equation for multi-species flow (\cite{Bird1960}, $\S$18.3).
In~\eref{heat:Eq} $\lambda$ is the effective heat conductivity of the mixture, $Q$ is the 
specific volumetric heat source due to plasma heating, $h_i$ is the variable part of the 
specific enthalpy per particle of the component $i$. The  $h_i$ is defined such that 
$h_i\left(T_{ref}\right) = 0$, where  $T_{ref}$ is some reference temperature chosen 
same for all components, and the full enthalpy is defined as 
$H_i\left(T\right) = H_i\left(T_{ref}\right) + h_i\left(T\right)$.
The constants $H_i\left(T_{ref}\right)$ which include enthalpy of formation are formally moved to the 
'chemical' heat source $S_h$ calculated with help of~\eref{diffusion:Eq} as:
\begin{equation}
\label{heat:source:Eq}
 S_h = {\rm div}{\left( \sum_i H_i\left(T_{ref}\right) \vec{\Gamma_i} \right) }  =  \sum_i H_i\left(T_{ref}\right) S_i
\end{equation}
To solve~\eref{heat:Eq} numerically it is translated into the standard conduction-convection form by
expressing $h_i$ as $h_i\left(T\right) = c^i_h\left(T\right)\cdot T$.
Radiation heat losses are not taken into account in~\eref{heat:Eq}.
Since only slow flows with mach numbers $M\ll$1 are considered kinetic energy of the bulk flow 
$\frac12 \rho v^2$ is omitted in~\eref{heat:source:Eq}, work of viscous stress is omitted as well.
Here $\rho = \sum_i m_in_i$, $m_i$ is the molecular weight of the component $i$. 

Equations~\eref{diffusion:Eq} and~\eref{heat:Eq} are solved assuming axial symmetry
in coordinates $\left(z,\;r\right)$, where $z$ is the coordinate along the axis and $r$ is the distance from the axis. 
The approximate model applied to find the bulk flow velocity $\vec{v}$ is based on two assumptions. 
First is the assumption of constant pressure $p=const$ which is justified, again, 
by small $M$. In all calculations which will be considered 
below the peak local Mach numbers of the flow are always $<$0.1. 
The second is the assumption of zero bulk radial velocity, $v_r$=0, that is, 
the mass transfer in the radial direction is neglected. This latter assumption has no 
formal justification, but it allows to greatly simplify the calculations.
In a channel with constant cross-section this condition implies local 
mass conservation, thus, the axial velocity $v_z$ is readily calculated as:
\begin{equation}
\label{flow:Eq}
\rho v_z = \Gamma_\rho\left(r\right) \quad \Rightarrow
\quad v_z = \frac{ \Gamma_\rho\left(r\right)}{ n \sum_i m_i X_i }
\end{equation}
Where $\Gamma_\rho\left(r\right)$ is determined by the boundary condition at the inlet. 
The total number density $n$ is calculated from the equation of state,~\sref{thermodynamic:data:S}.

The set of equations~\eref{diffusion:Eq},~\eref{heat:Eq},~\eref{flow:Eq} 
is solved numerically by a self-written finite-volume code. 
Some more details on implementation and numerics are given in~\ref{numerics:A}.

\subsection{Physico-chemical model}

\label{phys:chem:model:S}

The physico-chemical model as it is defined here comprises:
\begin{enumerate}
 \item The chemical mechanism: stoichiometric equations and reaction rate coefficients required to 
       calculate the source terms $S_i$ in~\eref{diffusion:Eq}
 \item The transport coefficients $D_{im}$ and $\lambda$ which enter~\eref{diffusion:Eq},~\eref{heat:Eq}
 \item Thermodynamic data
\end{enumerate}

\subsubsection{Chemical mechanism}

Five components are taken into account: CO$_2$, CO, O$_2$, O, C. 
The assumed chemical mechanism is summarized in~\tref{chemical:mechanism:T}. 
The mechanism consisting of the processes N1-N3 goes back to~\cite{Butylkin1979}, 
this same mechanism was applied in~\cite{denHarder2017,vandenBekerom2019}. 
The processes N4, N5 were added following the suggestion of~\cite{Wolf2020a} 
and because our own experience indicated that inclusion of the CO dissociation is 
mandatory to obtain solutions with realistic level of temperature. 

\Table{\label{chemical:mechanism:T} List of chemical processes.}
\br
N1 & CO$_2$ + M $\rightleftarrows$ CO + O + M \\
N2 & CO$_2$ + O $\rightleftarrows$ CO + O$_2$ \\
N3 & O + O + M  $\rightleftarrows$ O$_2$ + M \\
N4 & CO + M $\rightleftarrows$ C + O + M \\
N5 &  CO + O  $\rightleftarrows$  C + O$_2$ \\
\br
\endTable

The full set of chemical reactions included in the model is shown in~\tref{reactions:list:T}. 
The rate coefficients are calculated by applying the generalized Arrhenius equation:
\begin{equation}
\label{Arrhenius:Eq}
 R\left(T\right) = A T^\beta \exp{\left(-\frac{E_a}{T}\right)}
\end{equation}
The coefficients $A$, $\beta$ and $E_a$ are given in~\tref{reactions:list:T}.
An evaluation of the different sets of rate coefficients in a microwave CO$_2$ plasma conversion experiment 
was performed in~\cite{vandeSteeg2021} 
on the basis of spatially resolved Raman measurements of concentrations and comparison of the reconstructed and experimental conversion rate and input power. 
The conclusion of~\cite{vandeSteeg2021} was that the best results are obtained with the rate 
coefficients taken from the chemical mechanism GRI-Mech~3.0~\cite{GRIMech3}. 
This is a well known comprehensive chemical mechanism developed for modelling of methane combustion. 
Therefore, most of the rate coefficients used in the present work are taken from that data set. 
Rate coefficients of the process N3 with M=O,~O$_2$ and of the process N4 which are missing in~\cite{GRIMech3} 
are taken from data evaluations~\cite{Baulch1976,TsangHampson1986}. 

\begin{table}
\caption{\label{reactions:list:T} List of reactions.}
\begin{indented}
\item[] \begin{tabular}{@{}l|l|l|l|l|l|l}
\br
  &    & Stoichiometric equation & $A$ & $\beta$ & $E_a$ & Source \\
\mr
1 & N1 & CO$_2$ + CO$_2$ $\to$ CO + O + CO$_2$ & \multicolumn{4}{c}{$R_1=K_{eq}R_9$} \\
2 &    & CO$_2$ + CO $\to$  CO + O + CO & \multicolumn{4}{c}{$R_2=K_{eq}R_{10}$} \\
3 &    & CO$_2$ + O$_2$ $\to$ CO + O + O$_2$ & \multicolumn{4}{c}{$R_3=K_{eq}R_{11}$}  \\
\mr
4 & N2 & CO$_2$ + O  $\to$  CO + O$_2$ & \multicolumn{4}{c}{$R_4=K_{eq}R_{12}$}   \\
\mr
5 & N3 & O$_2$ + O$_2$ $\to$   O + O + O$_2$ & \multicolumn{4}{c}{$R_5=K_{eq}R_{13}$}  \\
6 &    & O$_2$ + O  $\to$  O + O + O &  5.8120$\cdot$10$^{-5}$ & -2.5 &  59380 & \cite{Baulch1976} \\
7 &    & O$_2$ + CO $\to$   O + O + CO & \multicolumn{4}{c}{$R_7=K_{eq}R_{15}$}  \\
8 &    & O$_2$ + CO$_2$ $\to$    O + O + CO$_2$ & \multicolumn{4}{c}{$R_8=K_{eq}R_{16}$} \\
\mr
9 & N1  & CO + O + CO$_2$ $\to$  CO$_2$ + CO$_2$ &  5.8101$\cdot$10$^{-45}$ & 0 & 1510.7 & \cite{GRIMech3} \\
10 &    & CO + O + CO $\to$  CO$_2$ + CO &  2.4900$\cdot$10$^{-45}$ & 0 & 1510.7 &   \cite{GRIMech3} \\
11 &    & CO + O + O$_2$  $\to$  CO$_2$ + O$_2$ & 9.9602$\cdot$10$^{-45}$ & 0 &  1510.7 &  \cite{GRIMech3}  \\
\mr
12 & N2  & CO + O$_2$ $\to$  CO$_2$ + O & 4.1514$\cdot$10$^{-18}$ & 0 & 24070 & \cite{GRIMech3} \\
\mr
13 & N3 & O + O + O$_2$ $\to$ O$_2$ + O$_2$ &  2.2$\cdot$10$^{-40}$ & -1.5 & 0 & \cite{TsangHampson1986} \\
14 &    & O + O + O $\to$ O$_2$ + O &  \multicolumn{4}{c}{$R_{14}=K_{eq}R_6$} \\
15 &    & O + O + CO $\to$ O$_2$ + CO &  5.7908$\cdot$10$^{-43}$ &  -1 & 0 & \cite{GRIMech3} \\
16 &    & O + O + CO$_2$ $\to$ O$_2$ + CO$_2$ & 1.1912$\cdot$10$^{-42}$ & -1 & 0 & \cite{GRIMech3} \\   
\mr
17 & N4 & CO + CO$_2$ $\to$ C + O + CO$_2$ & 1.4613 & -3.52 & 128700 &  \cite{Baulch1976} \\
18 &    & CO + CO $\to$ C + O + CO & 1.4613 & -3.52 & 128700 &  \cite{Baulch1976} \\
19 &    & CO + O $\to$ C + O + O & 6.8580$\cdot$10$^{-15}$ & 0 & 98025 &   \cite{Baulch1976} \\
20 &    & C + O + CO$_2$ $\to$ CO + CO$_2$ & \multicolumn{4}{c}{$R_{20}=K_{eq}R_{17}$} \\
21 &    & C + O + CO $\to$ CO + CO & \multicolumn{4}{c}{$R_{21}=K_{eq}R_{18}$}  \\
22 &    & C + O + O $\to$ CO + O & \multicolumn{4}{c}{$R_{22}=K_{eq}R_{19}$} \\
\mr
23 & N5 & CO + O $\to$ C + O$_2$ &\multicolumn{4}{c}{$R_{23}=K_{eq}R_{24}$} \\
24 &    & C + O$_2$ $\to$ CO + O &  9.6311$\cdot$10$^{-17}$ & 0 &  290.05 & \cite{GRIMech3} \\
\br
\end{tabular}
\item[] $A$, $\beta$,  $E_a$ are the parameters of~\eref{Arrhenius:Eq}; $E_a$ is in Kelvin;
the units of $A$ are chosen such that when $T$ is in Kelvin the $R\left(T\right)$ in~\eref{Arrhenius:Eq} 
is in m$^3$/s for reactions of 2nd order, and in m$^6$/s for reactions of 3rd order
\item[] For calculation of the equilibrium constants $K_{eq}$ see~\eref{Keq:Eq}
\end{indented}
\end{table}

Equilibrium factors $K_{eq}$ which appear in~\tref{reactions:list:T} are calculated by the formula known from 
thermodynamics (see e.g.~\cite{ReactingFlowBook2003}, Eq.~(9.43) there):
\begin{equation}
\label{Keq:Eq}
K_{eq} = \left(\frac{p_{ref}}{k_B T}\right)^{\sum_j \nu_j} \exp{ \left(-\frac{ \sum_j \nu_j G^{ref}_j\left(T\right) }{k_B T} \right) }
\end{equation}
where $\nu_i$ are the stochiometric factors of each chemical component in the corresponding process, 
by definition $\nu_j>0$ for products and  $\nu_j<0$ for reactants, the sums are taken  
over all components which take part in the given reaction; 
$G^{ref}_j\left(T\right)$ are Gibbs energies of the components at the reference pressure $p_{ref}$, 
$k_B$ is the Boltzmann constant.

\subsubsection{Transport coefficients}

\label{transport:coefficients:S}

The mixture averaged diffusion coefficient $D_{im}$ in~\eref{diffusion:Eq} is calculated 
by means of the expression derived in~\cite{CurtissHirschfelder1949}: 
\begin{equation}
\label{mixture:diffusion:Eq} 
D_{im} = \frac{ 1 - \frac{m_i n X_i }{\rho } }{ \sum_{j\ne j} \frac{X_i}{ \mathcal{D}_{ij} } }
\end{equation}
Here $\mathcal{D}_{ij}\left(T,p\right)$ 
is the binary diffusion coefficient calculated for the total pressure of the mixture $p$. 
According to~\cite{CurtissHirschfelder1949}~\eref{mixture:diffusion:Eq} 
gives correct approximation when $i$ is a trace impurity whose concentration is small. 
One can also easily show that this formula gives exact result for binary mixtures. 

Most of the coefficients $\mathcal{D}_{ij}$ used in the present work are
taken from~\cite{MarreroMason1972}, Equation (4.3-2), Table 13 there. 
For the pairs missing in~\cite{MarreroMason1972} the Fuller scaling is used: 
~\cite{PropertiesGasesLiquids5th}, Section 11-4, Equation~11-4.4 and Table~11-1 there.

The thermal conductivity of the mixture is calculated by applying the 
Wassilijewa Equation (see~\cite{PropertiesGasesLiquids5th}, Equation (10-6.1) there):
\begin{equation}
 \label{thermal:conductivity:eq}
 \lambda = \sum_i\frac{ X_i \lambda_i  }{ \sum_j  X_j \phi_{ij} } 
\end{equation}
where $\lambda_i\left(T\right)$ are the thermal conductivities of individual components, and $\phi_{ij}$ are defined as 
('Mason and Saxena Modification', Equation~(10-6.4) in~\cite{PropertiesGasesLiquids5th}):
\begin{equation}
\label{wilke:eq}
\phi_{ij} =  \frac{ \left[ 1 + \left(\frac{\mu_i}{\mu_j}\right)^{1/2}\left(\frac{m_j}{m_i}\right)^{1/4} \right]^2 }
{\left[ 8\left( 1 + \frac{m_i}{m_j} \right)\right]^{1/2} }
\end{equation}
where $\mu_i$ is the dynamic viscosity. In the calculations a simplified version of this 
formula is used obtained on assumption that for the dependency of $\mu$ on 
the molecular weight $m$ the scaling $\mu\sim\sqrt{m}$ holds. This scaling follows from the 
known theoretical expression for viscosity derived for spherical molecules without 
internal degrees of freedom (\cite{Hirschfelder1954}, Equation (8.2-10) there).
Then~\eref{wilke:eq} is reduced to 
\begin{equation}
\label{wilke:reduced:eq}
 \phi_{ij} = \sqrt{ \frac{2}{ 1 + \frac{m_i}{m_j}} }
\end{equation}
It is worth to note that there was no particular reason to use specifically the 
equations~\eref{thermal:conductivity:eq},~\eref{wilke:reduced:eq} for the  
calculation of $\lambda$. It is not unlikely that other more simple expressions which can be calculated faster, 
such as the formula suggested in~\cite{ReactingFlowBook2003}, Equation (12.119) there, will yield equally good results.

To calculate the coefficients $\lambda_i$ for CO$_2$, CO, O$_2$ the correlation formulas 
are used obtained by combining 
experimental data for low temperatures with theoretical extrapolations to higher temperatures. 
The $\lambda_i$ for CO$_2$ is taken from~\cite{Huber2016}, Equation~(3), Table~3 there. 
CO and N$_2$ are known to have very close thermal conductivities~\cite{MillatWakeham1989}. 
Therefore, as $\lambda_{CO}$ the thermal conductivity of N$_2$ is used, 
taken from~\cite{LemmonJacobsen2004}, Equations (5), (2), Tables I, II, IV there. 
The $\lambda_{O_2}$ is taken from the same reference~\cite{LemmonJacobsen2004}.
Thermal conductivity of O atoms is a fit to the results of calculations reported in~\cite{Holland1988}, Table~VI.
For C atoms $\lambda_i$ is calculated from the diffusion coefficient defined by
Fuller scaling applying the theoretical formula which connects $\lambda$ and the coefficient of 
self diffusion, see~\ref{kin:theory:A}.

More detailed information on the evaluation and cross-check of the literature data on $\mathcal{D}_{ij}$ 
and $\lambda_i$ is given in supplemental material. 

\subsubsection{Thermodynamic data}

\label{thermodynamic:data:S}

The thermodynamic parameters $h_i\left(T\right)$, $h_i\left(T_{ref}\right)$ in~\eref{heat:Eq}
as well as $G^{ref}_i\left(T\right)$ in~\eref{Keq:Eq} 
are all taken from~\cite{McBride2002}; $T_{ref}$=0~K, $p_{ref}$=10$^5$~Pa. 
The equation of state is that of ideal gas $p=nk_BT$.

\section{Comparison with experiment}

\label{comparison:with:experiment:S}

\subsection{Description of the experiment and its computational model}

\label{reference:model:S}

\begin{figure}
\center
\subfloat[]{\label{experiment:sketch:fig} 
\includegraphics[width=8cm]{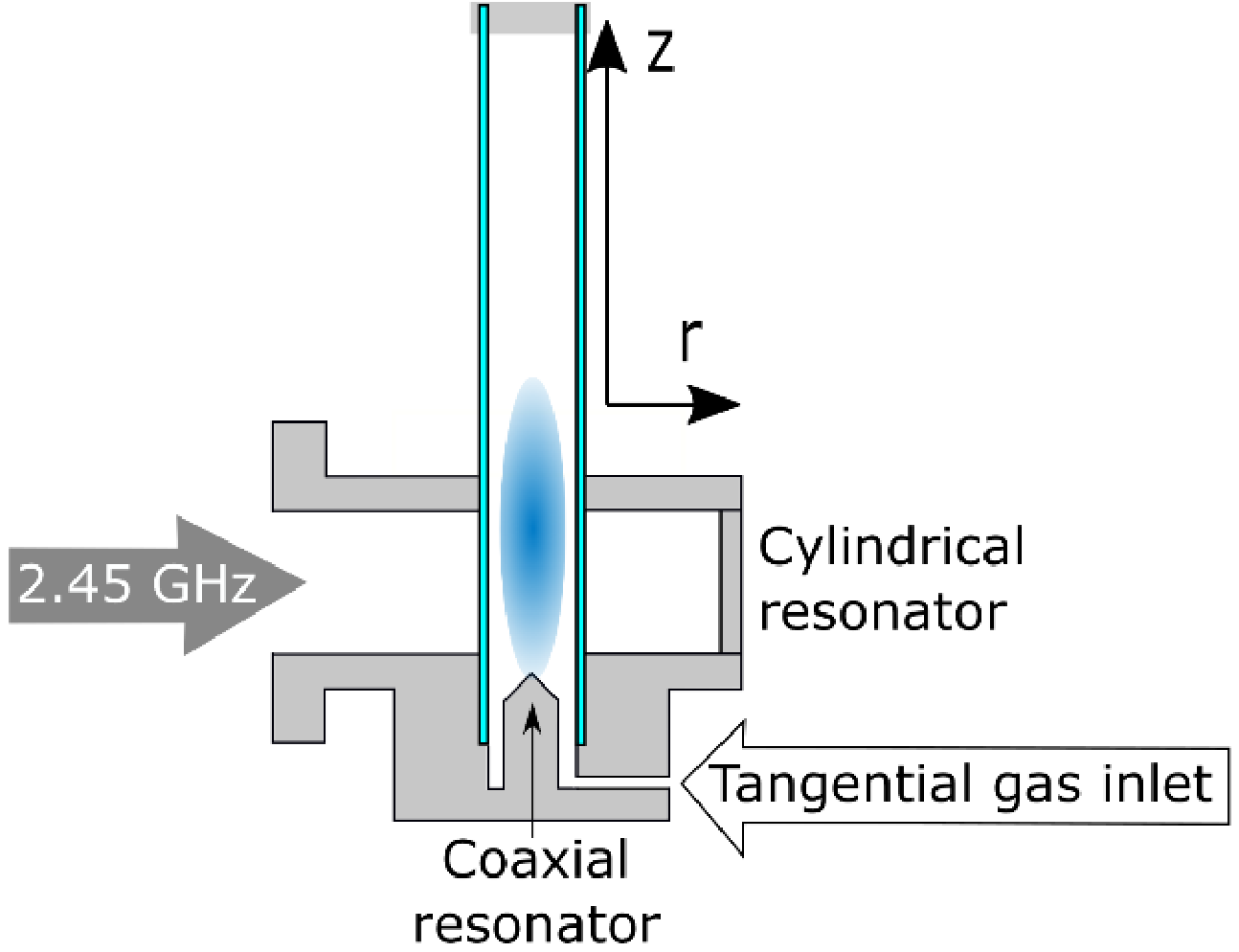} }
\subfloat[]{\label{model:sketch:fig} 
\includegraphics[width=3.5cm]{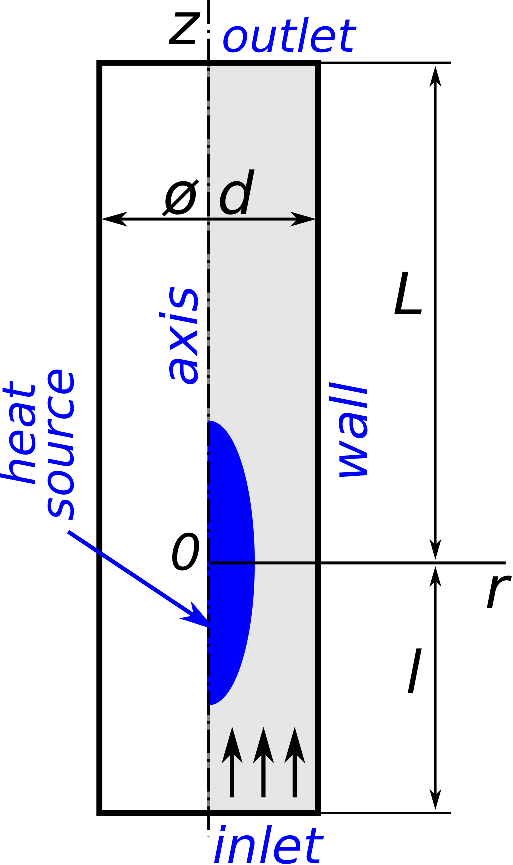} }
\caption{(a) Sketch of the experimental device; (b) its computational model}
\end{figure}

The experimental set-up will be only very briefly described here since its detailed description can be 
found in~\cite{DIsaThesis2021} and in the open access publication F.~A.~D'Isa~et~al.~2020~\cite{DIsa2020}. 
The discharge is ignited in the quartz tube which crosses a resonator where 2.45~GHz wave is excited by a magnetron generator, 
see~\fref{experiment:sketch:fig}. 
The resonator design of Stuttgart University is applied consisting of the 
main cylindrical resonator with $E_{010}$ wave ($E$-field is directed along the quartz tube axis) for continuous operation and 
the additional coaxial resonator serving for ignition of plasma~\cite{Leins2007,Leins2009}. 
The working gas is injected through four tangential gas inlets. 
The discharge tube is followed by the heat exchanger connected to a vacuum pump. 
The conversion rate is obtained by measuring the mixture composition with a mass spectrometer~\cite{Hecomovic2020}. 
Optical access is available from the window on the top and through the side slit in the cylindrical resonator. 
Optical emission spectroscopy is applied to measure the spatial distribution of the plasma radiation
as well as for reconstruction of the gas temperature~\cite{Carbone2020}.

The computational model simplifies the real geometry by a straight cylindrical channel,~\fref{model:sketch:fig}.
In particular, the tip of the coaxial resonator at the bottom is not taken into account.  
The origin of the coordinate system $\left(z,\, r\right)$ corresponds to the center 
of the heat source where it assumes its peak value, equation~\eref{Q:shape:eq} below. 
This point is located on the discharge tube axis close to the middle of the cylindrical resonator.
The internal diameter of the channel $d$=26~mm, the length $l$ from the gas inlet to the origin 
is set to $l=2l_{pl}$, where $l_{pl}$ is the estimated plasma length (see below). 
The length $L$ from the origin to the end of the computational domain 
is set to $L=$1~m (in some cases to  $L=$1.5~m). 
This choice ensures that at that location in all the simulation 
runs considered the gas temperature is guaranteed below 1500~K, 
the molar fraction of oxygen atoms is always $<$2$\cdot$10$^{-5}$, and that 
to this point all the chemical transformations have been completed. 

The boundary conditions of equations~\eref{diffusion:Eq},~\eref{heat:Eq} are described in~\tref{boundary:conditions:T}. 
Note that constant $\Gamma_n$ corresponds to uniform spatial distribution of the CO$_2$ flux through the inlet. 
$\Gamma_\rho$ in~\eref{flow:Eq} is $\Gamma_n$ multiplied by molecular weight of CO$_2$.
In the experiment the wall temperature $T_w$ is known to be elevated, but its exact value was not measured. 
Thus, here room temperature is specified everywhere on the wall on implicit assumption that the 
solution must not be sensitive with respect to that boundary condition.
This assumption was checked in one modelling case (0.9~bar, 0.9~kW, 10~slm, see~\sref{results:S} below)  
where it was shown that increasing $T_w$ to 600~K does not change the calculated 
conversion rate.

\begin{table}
\caption{\label{boundary:conditions:T} Boundary conditions for the computational domain of~\fref{model:sketch:fig}.}
\begin{indented}
\item[] \begin{tabular}{@{}lll}
\br
 Boundary & \Eref{diffusion:Eq} & \Eref{heat:Eq} \\
\mr
 Inlet  & $i=CO_2$: $\Gamma_{iz}=\Gamma_n$ & $T$=300~K \\ 
        & $i\ne CO_2$: $\Gamma_{iz}=0$   &           \\
 Wall   & $\Gamma_{ir}$=0 & $T$=300~K \\ 
 Axis   & $\partial X_i/\partial r=0$  & $\partial T/\partial r=0$ \\ 
 Outlet & $\partial X_i/\partial z=0$  & $\partial T/\partial z=0$ \\ 
\br
\end{tabular}
\item[] $\Gamma_n$ is the total flow rate divided by $\pi d^2/4$ 
\end{indented}
\end{table}

The spatial shape of the power source is specified on the basis of the measured distribution of the plasma radiation intensity. 
The basic thought behind this method is that since the power is introduced into the gas via free 
electrons the specific volumetric input power should be approximately proportional to the electron density $n_e$. 
At the same time, the line radiation intensity is proportional to $n_e$ as well. Therefore, 
the profile of the radiation intensity should roughly reflect the profile of the specific input power. 
More thorough analysis~\cite{Wolf2019,Viegas2021} has confirmed that the profile of the 
777~nm O-atoms line radiation is indeed representative - to a certain degree of accuracy - 
of at least the radial power deposition profile. 
In F.~A.~D'Isa et al.~\cite{DIsa2020,DIsaThesis2021} not this particular line was used, but the total radiation which 
in contracted discharge mode is shown to be mainly due to C$_2$ bands. 
Nevertheless, extra measurements done with filter confirmed that the distribution of the total radiation is 
the same as that of the oxygen only~\cite{DIsa2020}.

The measured radial radiation profile has approximately Gaussian shape. 
Therefore, in the calculations the spatial distribution of the specific input power 
is approximated by the following function:
\begin{equation}
\label{Q:shape:eq}
Q\left(z,r\right) = c\cdot  \exp{\left( -\alpha_z z^2 \right)} \exp{\left( -\alpha_r r^2 \right)}
\end{equation}
where the constant $c$ is adjusted such that the integral of $Q\left(z,r\right)$ over the 
computational volume equals to the prescribed total input power. 
Whereas the use of Gaussian function for the radial profile 
is well justified this is less certain for the axial profile.
Its use in the present paper goes back to the work~\cite{Groen2019} where $Q\left(z,r\right)$
was reconstructed by applying the impedance matching technique and both $z$ and $r$ profiles were 
fitted by Gaussian functions. However, there are indications that for the experiments D'Isa et al.~\cite{DIsa2020,DIsaThesis2021}
a shape less peaked in the middle could be more appropriate. 
Therefore, while in the reference calculations always~\eref{Q:shape:eq} is applied several cases were repeated with flat $z$-profile 
to demonstrate that the final solution is not strongly affected by this assumption, see~\sref{sensitivity:S}. 

Parameters $\alpha_r$, $\alpha_z$ in~\eref{Q:shape:eq} are reconstructed from the 'plasma diameter' $d_{pl}$ and 'plasma length' $l_{pl}$ 
reported in~\cite{DIsa2020}, Figure~9 and Figure~11 there respectively. 
The plasma radius $d_{pl}/2$ is defined in~\cite{DIsa2020,DIsaThesis2021} as the radius where  
the radiation intensity is 15~\% of its maximum. Hence, $\alpha_r$ is calculated by applying the following rule:
\begin{equation}
\exp{\left( -\alpha_r \left(d_{pl}/2\right)^2 \right)} = 0.15\quad\Rightarrow\quad
\alpha_r = \frac{4\cdot 1.9}{d^2_{pl}}
\end{equation}
Exactly same rule is used to calculate $\alpha_z$ with $d_{pl}$ replaced by $l_{pl}$. 

It is known that the procedure of determining the characteristic dimensions of the power deposition 
profile from the plasma optical size is not accurate.  
A specific example of the error which may be introduced by this method 
can be found in~\cite{Viegas2021}. There it is shown that at pressures 
$<$150~mbar the power deposition radius is by a factor 1.6 larger than the optical plasma radius.
Therefore, evaluating the effect of increasing or reducing $d_{pl}$ and $l_{pl}$ 
is always part of the computational analysis when the experimental $Q\left(z,r\right)$ is used. 
The impact of this variation will be discussed in~\sref{sensitivity:S} below. 

\subsection{Comparing the reference model with the experiment}

\label{results:S}

\begin{figure}
\center
\subfloat[]{\label{09bar:Tmax:fig} 
\includegraphics[width=8cm]{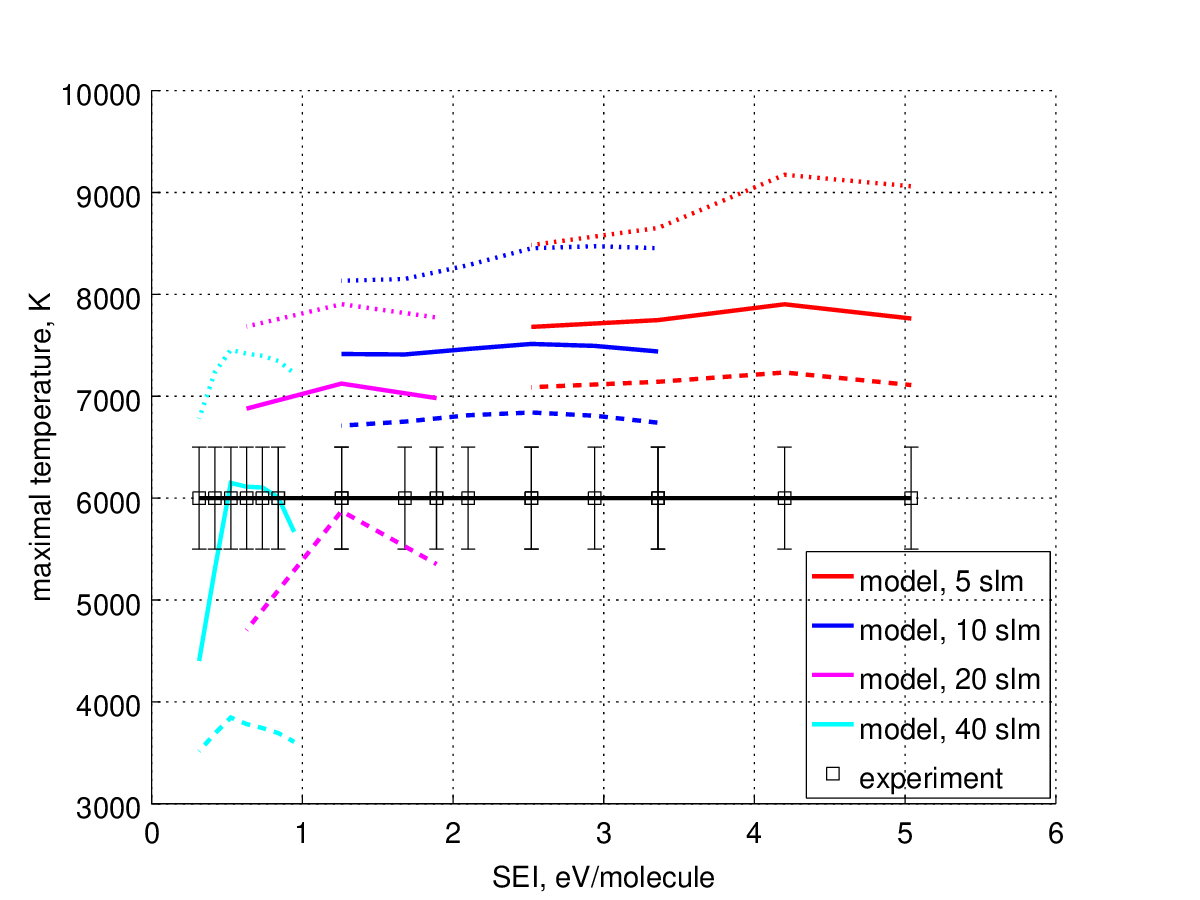} }
\subfloat[]{\label{09bar:X:fig} 
\includegraphics[width=8cm]{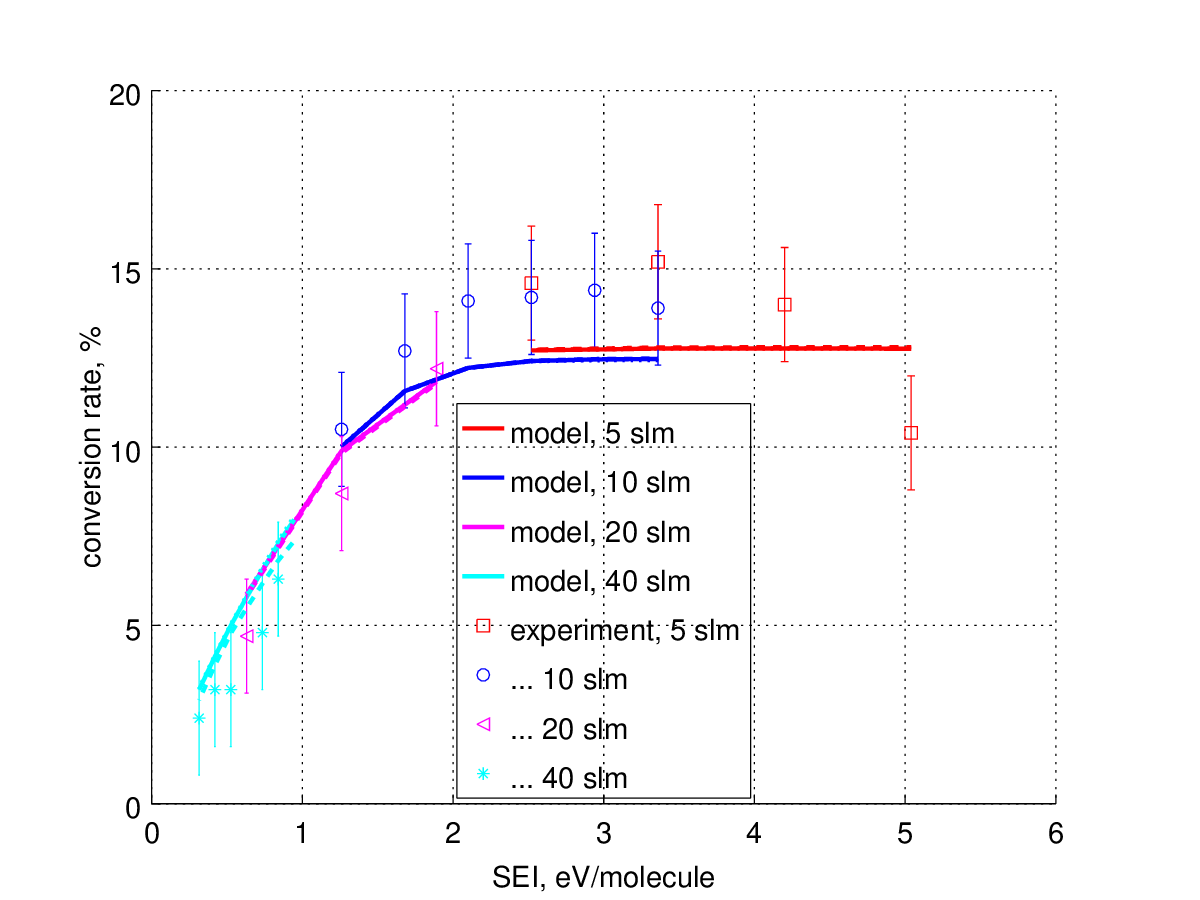} } \\
\subfloat[]{\label{09bar:eta:fig} 
\includegraphics[width=8cm]{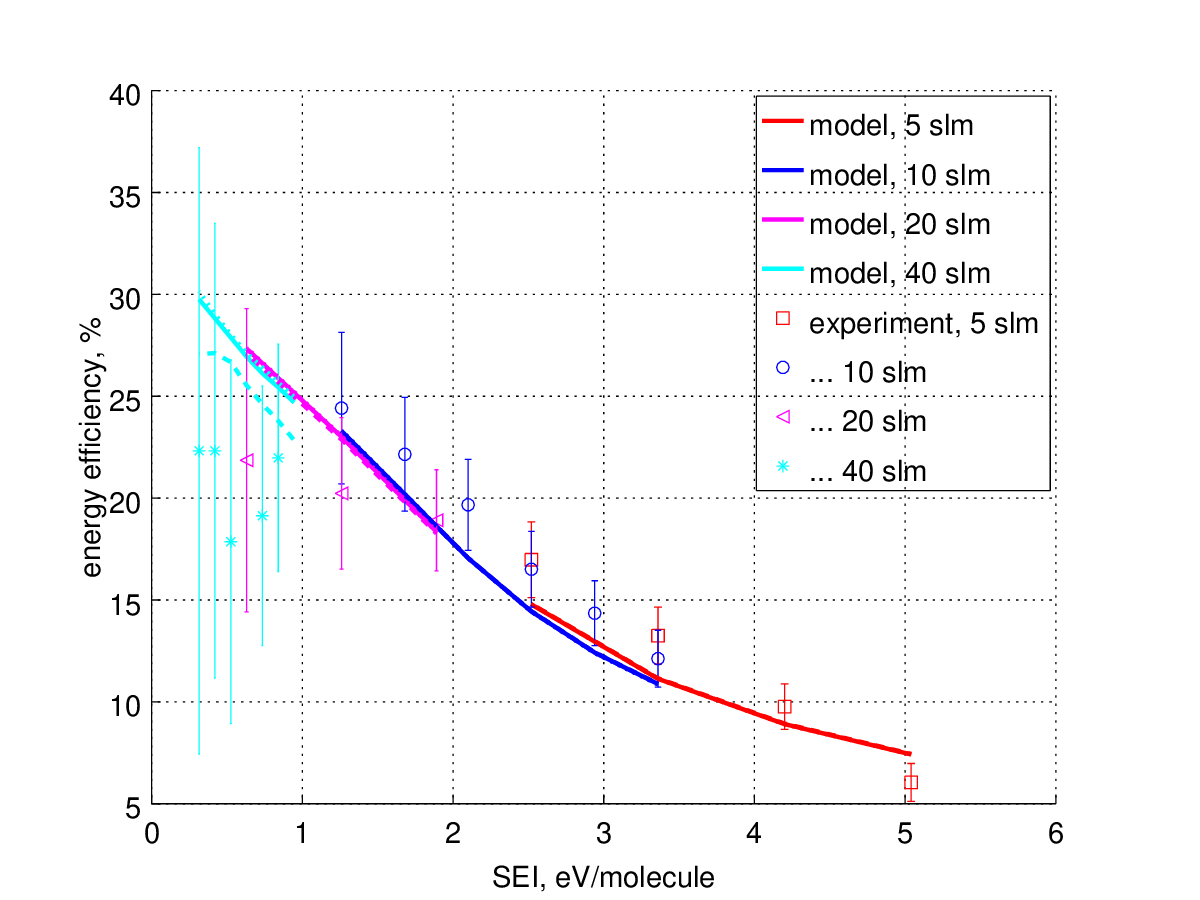} }
\caption{Comparison of the model with the experiments~\cite{DIsa2020,DIsaThesis2021}
for pressure 0.9~bar and different flow  rates. SEI is defined by~\eref{SEI:definition:eq},
'conversion rate' $\chi$ is defined by~\eref{conversion:definition:eq} and 
'energy efficiency' $\eta$ by~\eref{efficiency:definition:eq}. 
The dashed lines are obtained with the nominal value of $d_{pl}$ increased by a factor 1.5, 
the dotted lines with $d_{pl}$ decreased by a factor 1.5, see~\sref{sensitivity:S} }
\label{09bar:results:fig} 
\end{figure}

The results of comparison of the calculations performed with the reference model described in~\sref{reference:model:S}  
above with the experiment F.~A.~D'Isa~et~al.~\cite{DIsa2020,DIsaThesis2021} at quasi-atmospheric pressure 0.9~bar are presented 
in~\fref{09bar:results:fig}. The peak temperature $T_{max}$, conversion rate $\chi$ and energy efficiency $\eta$ 
are plotted as functions of the Specific Energy Input per molecule SEI. This latter is defined as: 
\begin{equation}
 \label{SEI:definition:eq}
 \mbox{SEI [eV/molecule]} = 14\frac{\mbox{input power [kW]} }{ \mbox{flow rate [slm]} }
\end{equation}
where 'input power' is the power coupled into plasma. Measurements of the rotational temperatures~\cite{Carbone2020} 
have shown that in contracted mode in a wide range of experimental parameters $T_{max}$ always acquires 
similar values. Therefore, for all experiments this value is set here to $T_{max}=$6000$\pm$500~K~\cite{DIsa2020,Carbone2020}.
The conversion rate is defined as:
\begin{equation}
 \label{conversion:definition:eq}
  \chi  = \frac{ \mbox{CO outflux} }{ \mbox{flow rate} } =  1 -  \frac{ \mbox{CO$_2$ outflux} }{ \mbox{flow rate} }
\end{equation}
where 'CO outflux' in the model is the total flux of CO molecules through outlet (see~\fref{model:sketch:fig}), 
and in the experiment this is the flux obtained by means of the mass spectrometer measurements after the heat exchanger. 
The error bar of the measurements is set to $\pm$1.6~\% according to~\cite{Hecomovic2020}. 
The energy efficiency is connected to $\chi$ by the equation:
\begin{equation}
 \label{efficiency:definition:eq}
  \eta = \frac{\Delta H_f }{\mbox{SEI}}\chi
\end{equation}
where $\Delta H_f$=2.93~eV is the net enthalpy change in the chemical transformation CO$_2\to$CO+$\frac12$O$_2$ at 
$T$=298.15~K~\cite{McBride2002}.

The temperature $T_{max}$,~\fref{09bar:Tmax:fig}, is not very well reproduced by the model. 
In most cases the (reference) model tends to 
overestimate $T_{max}$ by 20..30~\%. In the case with the highest flow rate 40~slm, however,
the temperature may be even underestimated. 
Despite that, as one cane see in figures~\ref{09bar:X:fig},~\ref{09bar:eta:fig} 
the $\chi$ and $\eta$ are in a good agreement with the experiment.
The conversion rate is mostly within the experimental error bars, and the trend observed in the experiment is reproduced 
- except for the largest SEI (lowest flow rate). 
The reduction of $\chi$ observed when SEI is increased above 3~eV is not reproduced 
by the model which yields saturation of $\chi$ in this SEI range.  
Opposite to that, for $\eta$ the largest deviation can be seen at SEI$<$1~eV where the experimental 
data indicate nearly saturation with decreased SEI whereas in the model $\eta$ still increases, 
although the values are formally within experimental error bars. 

%
%

\begin{figure}
\center
\subfloat[]{\label{05bar:Tmax:fig} 
\includegraphics[width=5cm]{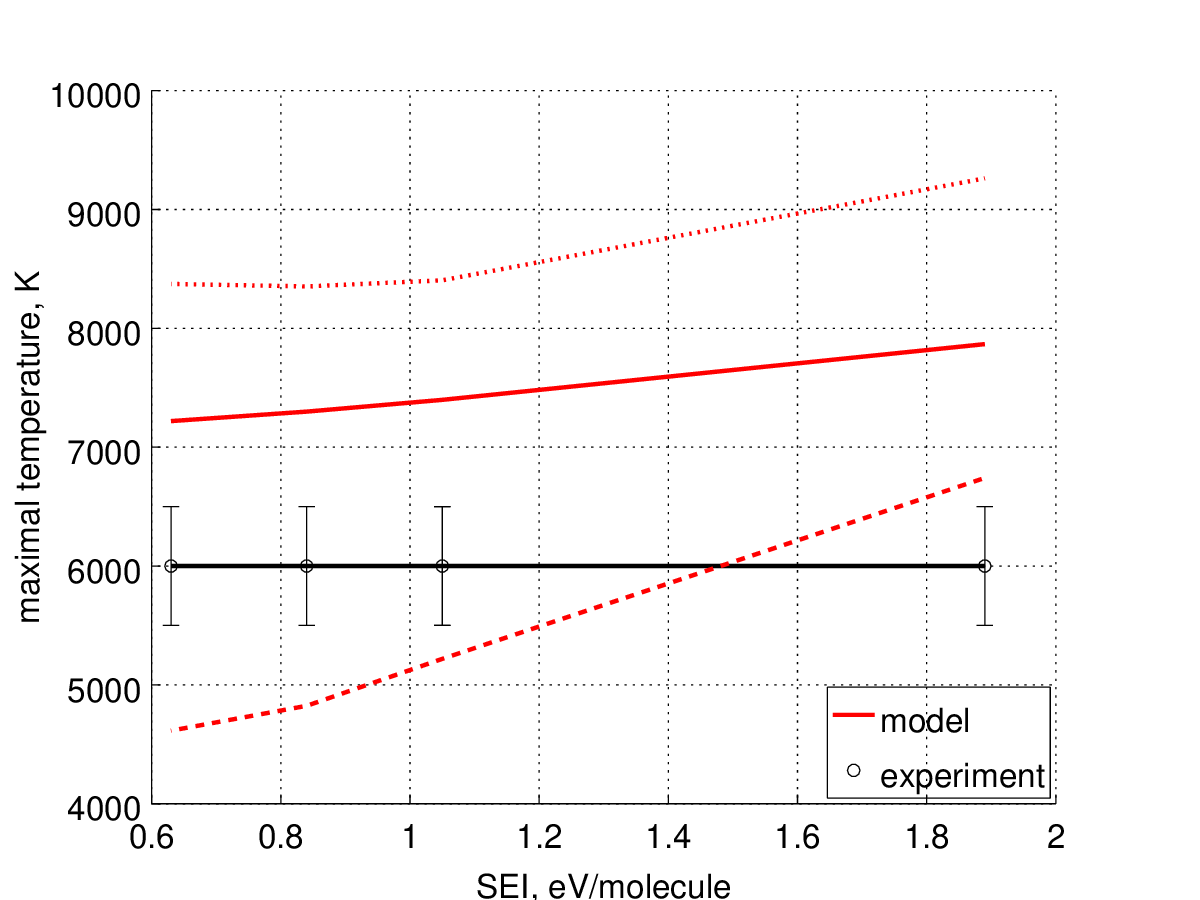} }
\subfloat[]{
\includegraphics[width=5cm]{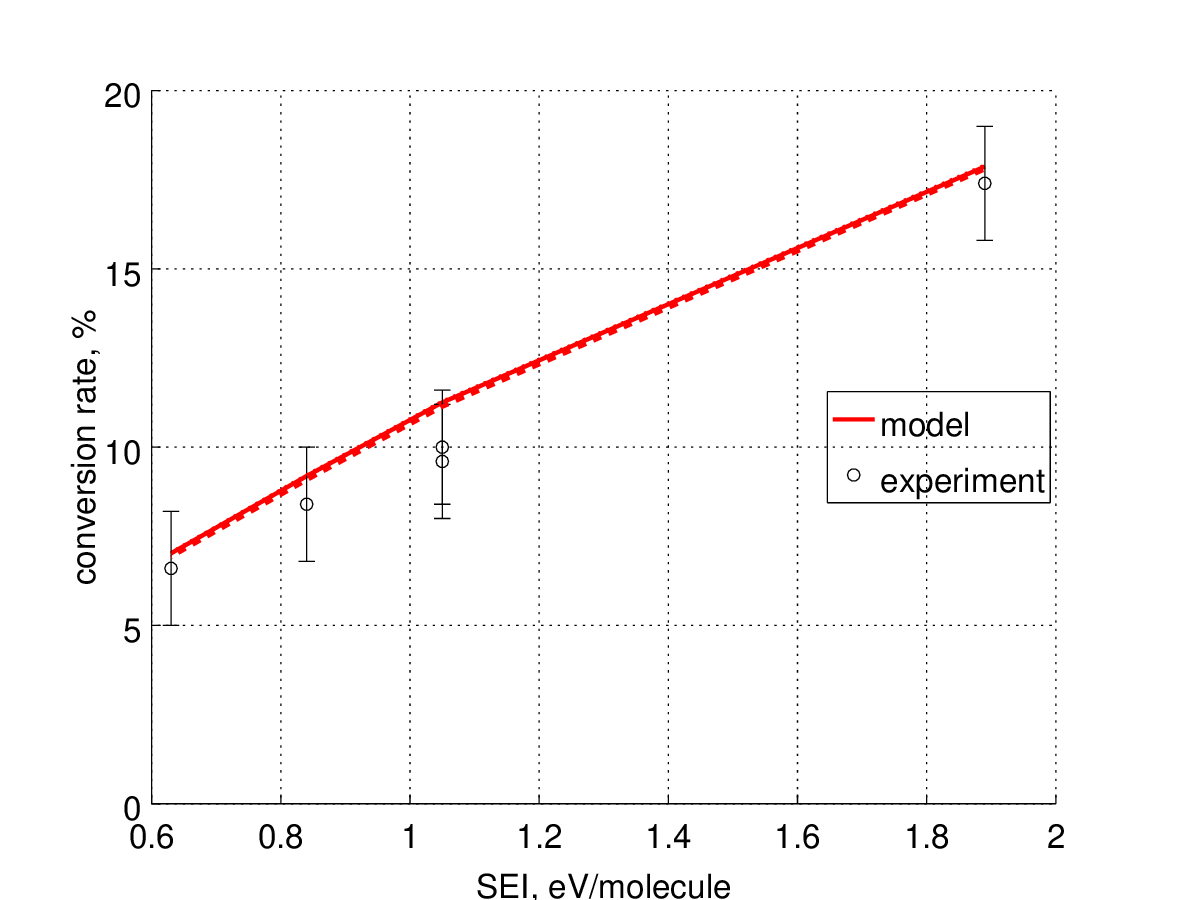} }
\subfloat[]{
\includegraphics[width=5cm]{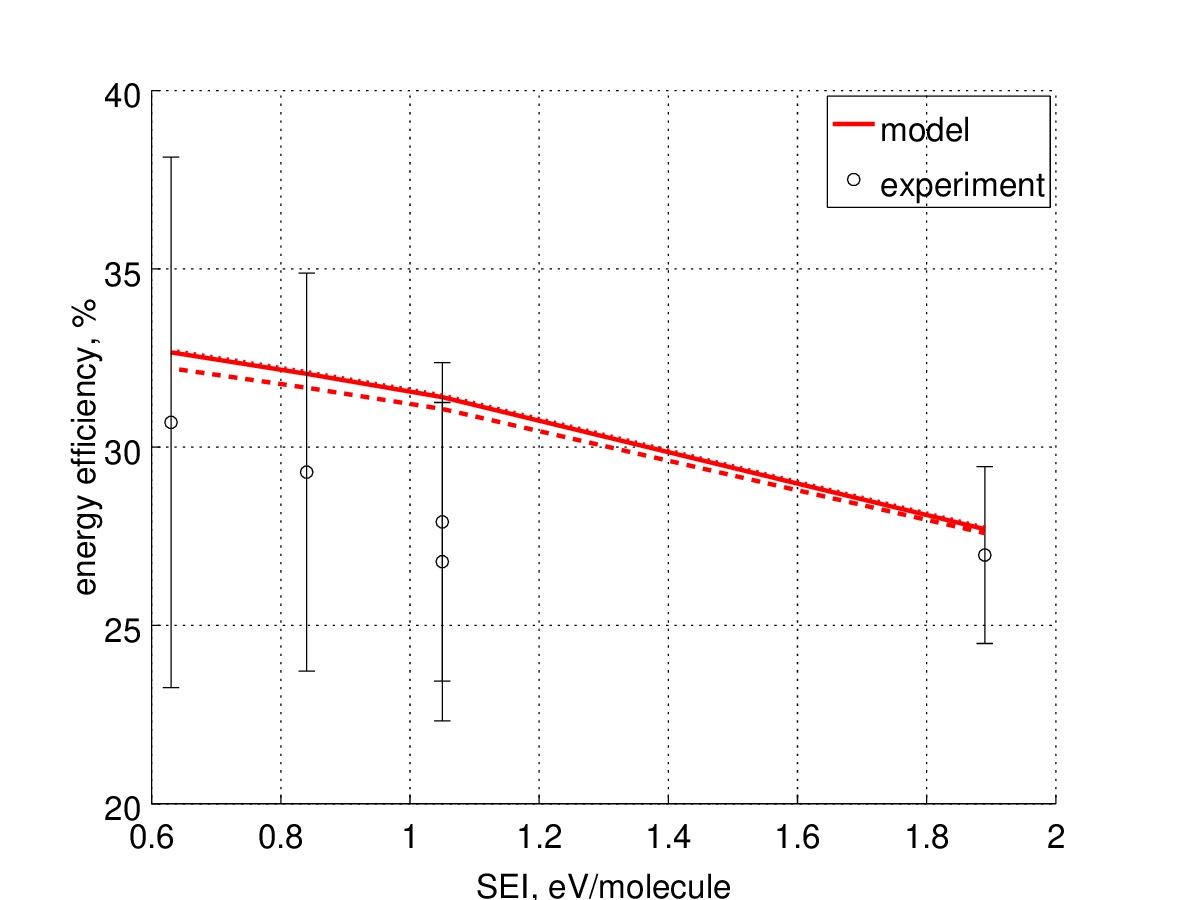} }
\caption{Comparison of the model with the experiments~\cite{DIsa2020,DIsaThesis2021}
for pressure 0.5~bar, flow rate 20~slm. SEI is defined by~\eref{SEI:definition:eq}, 
'conversion rate' $\chi$ is defined by~\eref{conversion:definition:eq} and 
'energy efficiency' $\eta$ by~\eref{efficiency:definition:eq}. 
The dashed lines are obtained with the nominal value of $d_{pl}$ increased by a factor 1.5, 
the dotted lines with $d_{pl}$ decreased by a factor 1.5, see~\sref{sensitivity:S} }
\label{05bar:results:fig} 
\end{figure}

\begin{figure}
\center
\subfloat[]{\label{02bar:Tmax:fig} 
\includegraphics[width=5cm]{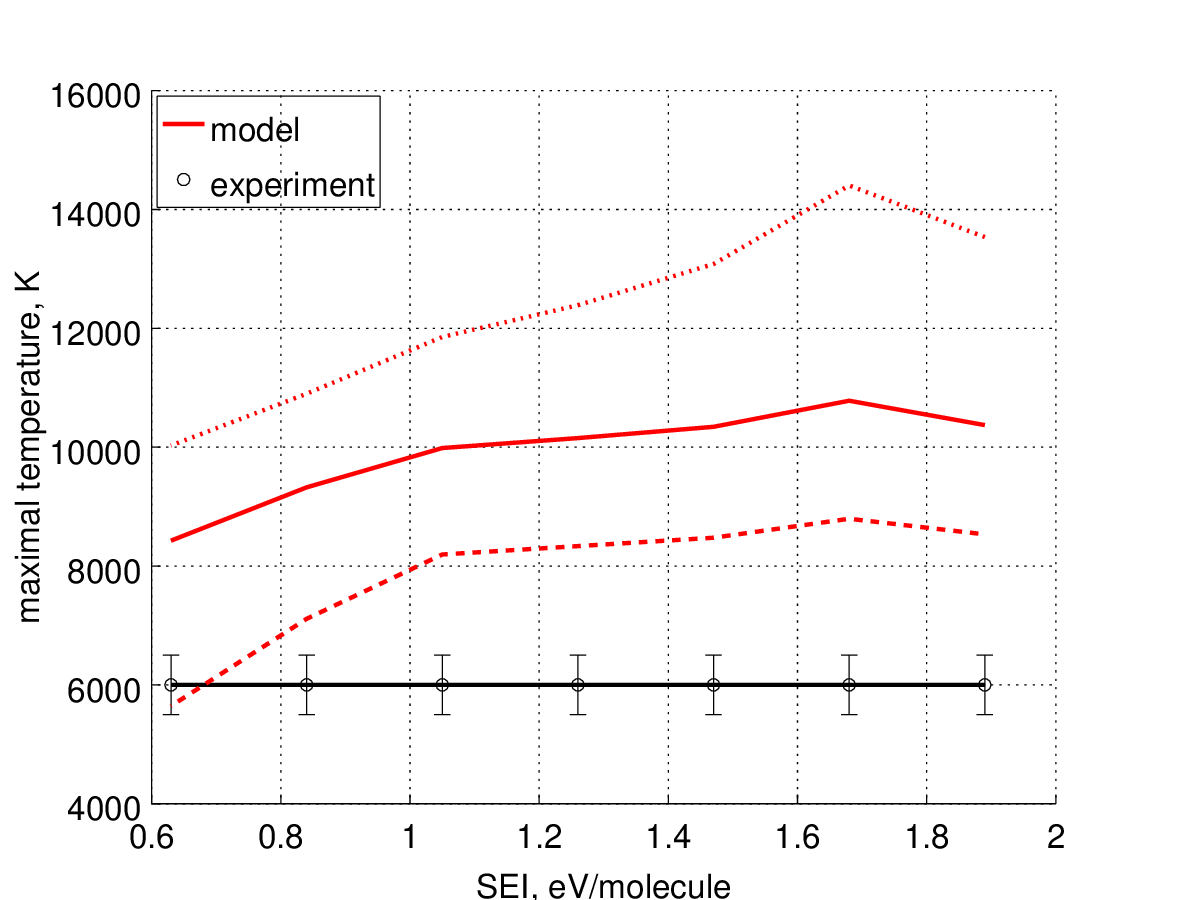} }
\subfloat[]{\label{02bar:X:fig} 
\includegraphics[width=5cm]{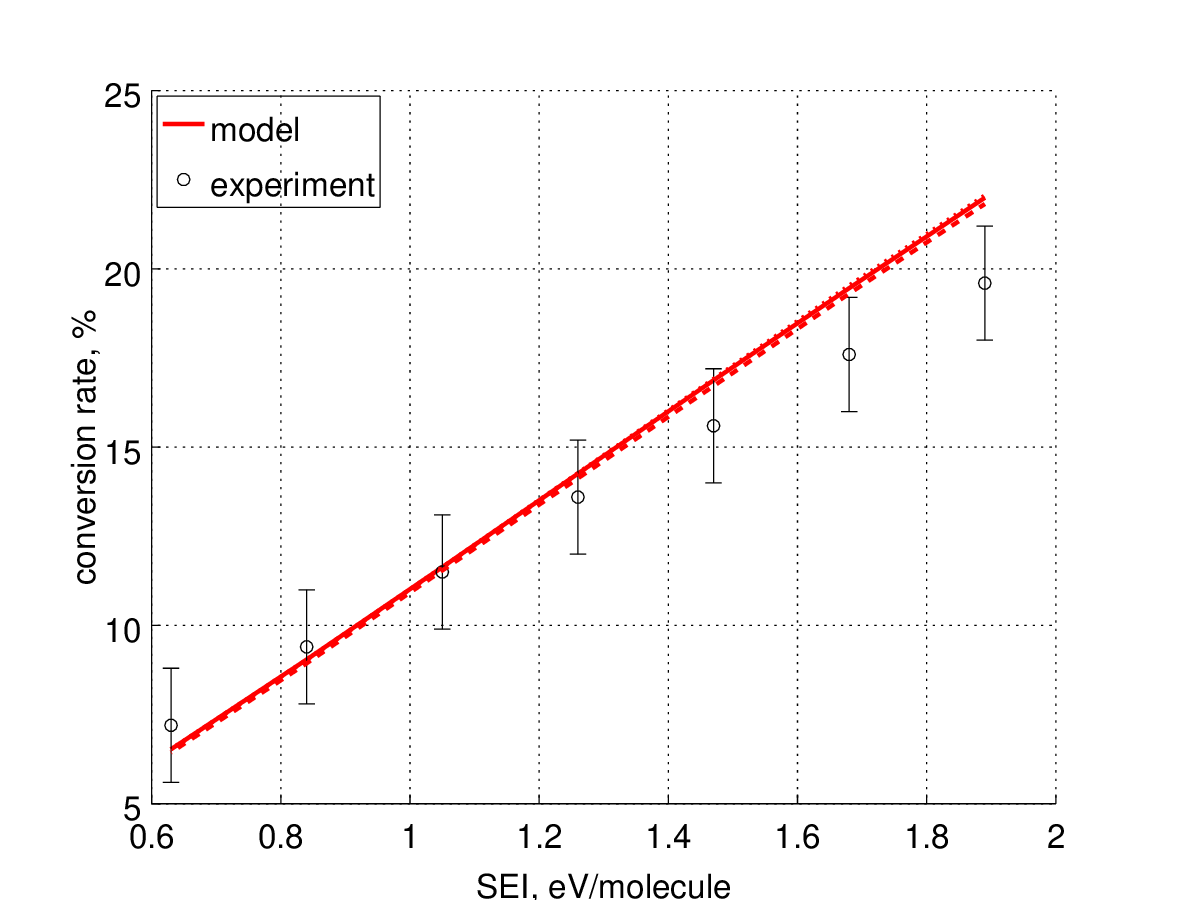} }
\subfloat[]{\label{02bar:eta:fig} 
\includegraphics[width=5cm]{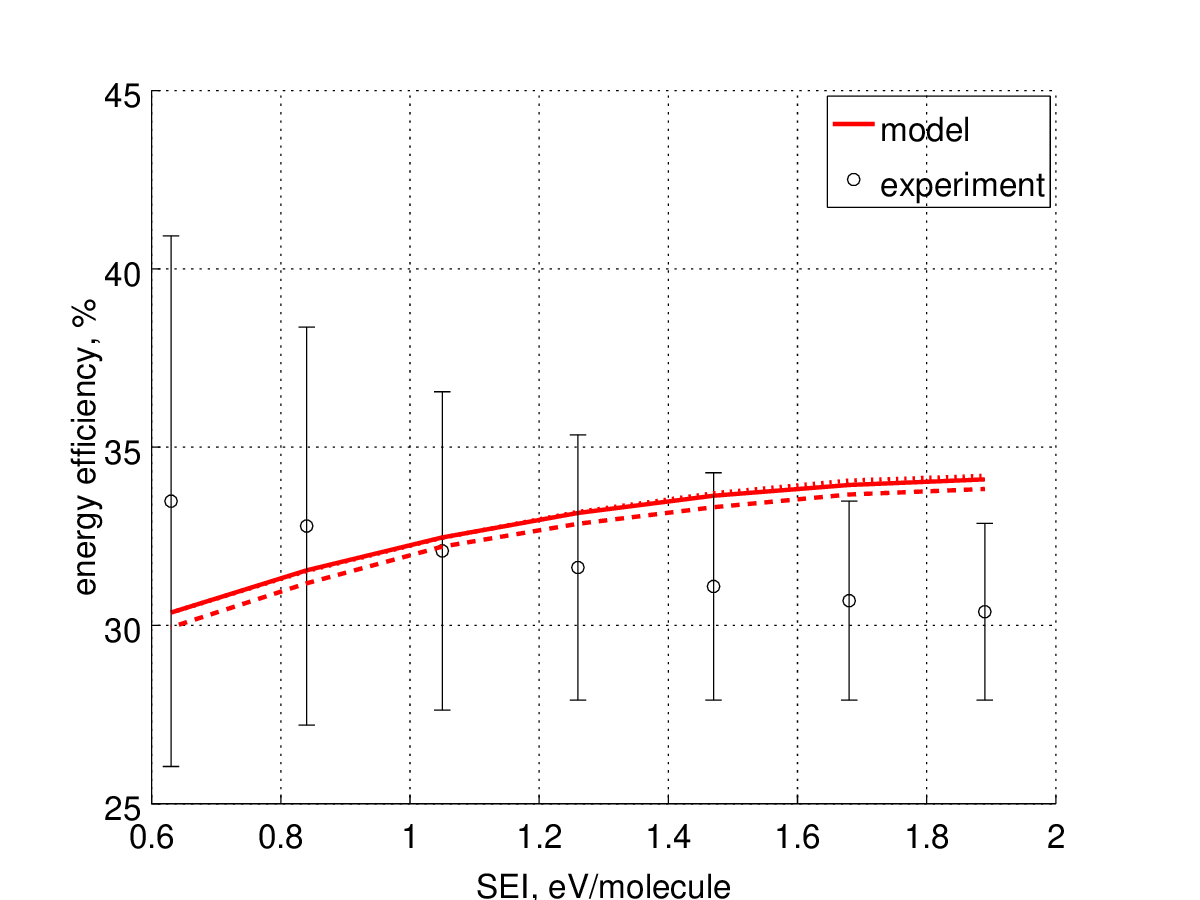} }
\caption{Comparison of the model with the experiments~\cite{DIsa2020,DIsaThesis2021}
for pressure 0.2~bar, flow rate 20~slm. SEI is defined by~\eref{SEI:definition:eq}, 
'conversion rate' $\chi$ is defined by~\eref{conversion:definition:eq} and 
'energy efficiency' $\eta$ by~\eref{efficiency:definition:eq}. 
The dashed lines are obtained with the nominal value of $d_{pl}$ increased by a factor 1.5, 
the dotted lines with $d_{pl}$ decreased by a factor 1.5, see~\sref{sensitivity:S} }
\label{02bar:results:fig} 
\end{figure}

The results for pressure 0.5~bar are similar to those for  0.9~bar,~\fref{05bar:results:fig}.  
The situation is different for the lowest investigated pressure 0.2~bar,~\fref{02bar:results:fig}.
In this case the model overestimates the temperature by up to a factor two,~\fref{02bar:Tmax:fig}. 
To the extent that the modelling results are physically inconsistent. 
At temperatures above 8000~K even in the state of local thermodynamic equilibrium the 
plasma cannot be weekly ionized anymore~\cite{Marieu2007} which contradicts the model assumptions. 
Surprisingly, despite this drastic mismatch the conversion rate and energy efficiency are 
still very well reproduced by the model,~\fref{02bar:X:fig},~\ref{02bar:eta:fig}. 
The main disagreement is that in the experiment $\eta$ is decreased with increased SEI, 
and in the model this trend goes into opposite direction. Although in both cases the trend is weak.

\begin{figure}
\center
\subfloat[]{\label{temperature:rad:fig} 
\includegraphics[width=7cm]{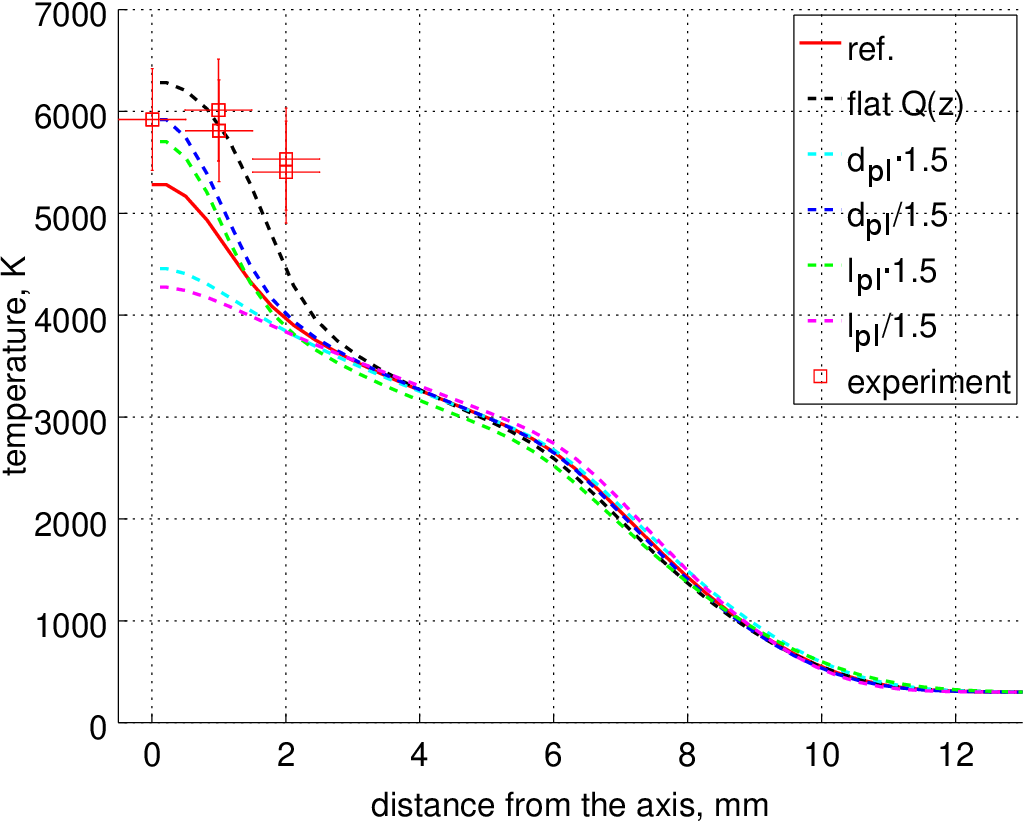} }
\subfloat[]{
\includegraphics[width=7cm]{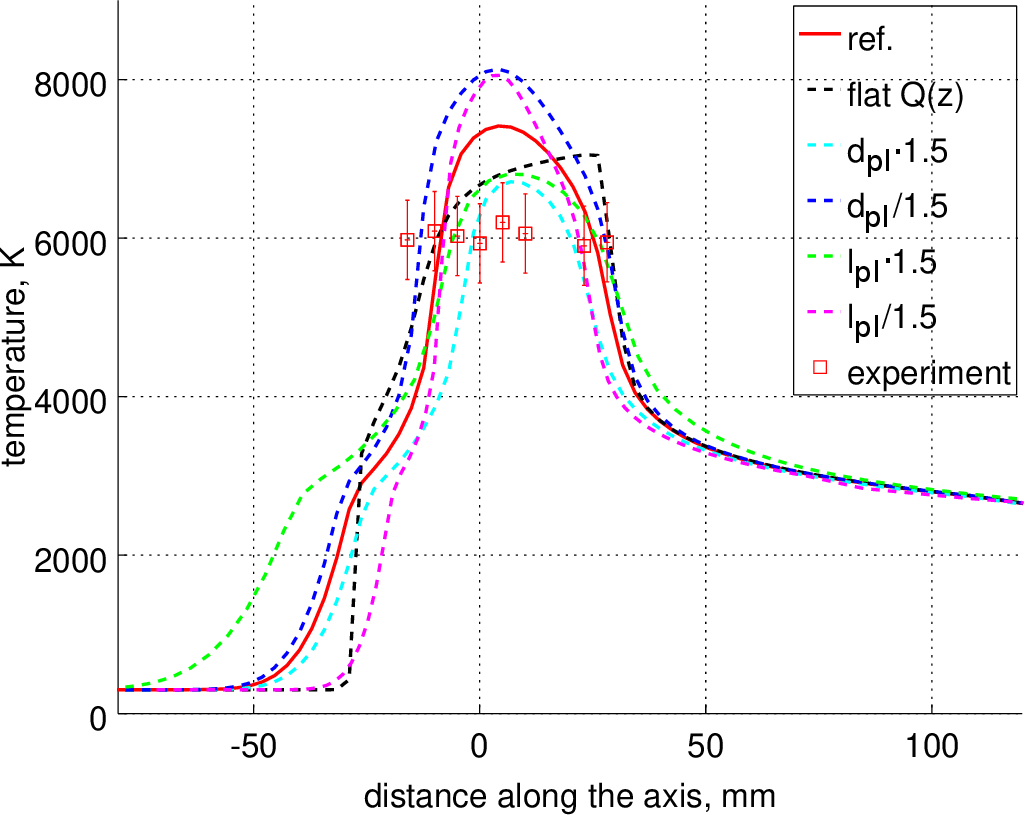} }
\caption{Spatial distribution of the gas temperature in the case 0.9~bar, 0.9~kW, 10~slm (same case as in Figure 6 in~\cite{DIsa2020}). 
         (a) Is the radial profile '58 mm above the resonator bottom'; (b) is the profile along the axis, $r=0$. 
         Solid lines is the reference model of~\sref{reference:model:S}, 
         dashed lines correspond to different cases listed in~\tref{sensitivity:Tmax:T} }
\label{temperature:fig} 
\end{figure}

The calculated profiles of the gas temperature are compared with measurements in~\fref{temperature:fig} 
which corresponds to the exactly same experimental case as figure~6 in~\cite{DIsa2020}. 
The error bars of the measured temperature are set to $\pm$500K, 
the error bars of the spatial coordinates in the measurements are set to $\pm$0.5~mm according to~\cite{DIsa2020}.
The radial profile corresponds to the location '58 mm at high above the resonator bottom'~\cite{DIsa2020}. 
To translate this position into the model coordinates (shown in~\fref{model:sketch:fig}) it was assumed 
that the center of the heat source is located 30~mm above the resonator bottom, as can be estimated approximately from Figure 10d in~\cite{DIsa2020}. 
One can see in~\fref{temperature:fig} that not only the absolute values of temperatures are different, 
but also that both radial an axial profiles in the model are peaked whereas in the experiment they are nearly flat.

%
%

\section{Discussion}

\subsection{Sensitivity assessment}

\label{sensitivity:S}

The sensitivity of the solution has been investigated for 6 selected cases. 
They are listed in~\tref{sensitivity:Tmax:T} which presents the variation of the maximum temperature $T_{max}$. 
In this table the column 'ref.' is the solution obtained with the reference model of~\sref{reference:model:S}.  
Next columns show the results of varying the spatial shape of the heat source $Q\left(z,r\right)$. 
The column 'flat Q(z)' stands for the heat source described by the function 
$$
Q\left(z,r\right) = 
\left\{
\begin{array}{lc}
c\cdot \exp{\left( -\alpha_r r^2 \right)}, & -l_{pl}/2 \leq z \leq l_{pl}/2 \\
0, & \mbox{otherwise} 
\end{array}\right.
$$ 
instead of~\eref{Q:shape:eq}. Other columns show the effect of increasing and decreasing the assumed plasma
diameter $d_{pl}$ and plasma length $l_{pl}$ by a factor 1.5 compared to their nominal values taken from~\cite{DIsa2020,DIsaThesis2021}.

Whereas the effect of switching to the flat $Q\left(z\right)$-profile on $T_{max}$ is visible but limited, 
the effect of the variation of $d_{pl}$, $l_{pl}$ is large.  
Its magnitude is further illustrated in figures~\ref{09bar:Tmax:fig}-\ref{02bar:Tmax:fig} 
where the solutions obtained with increased and reduced $d_{pl}$ are plotted as dashed and dotted lines respectively.  
At lower pressures the effect of the variation of $d_{pl}$ on $T_{max}$ is notably stronger than at the quasi-atmospheric pressure.
\Fref{temperature:fig} presents an example of the modification of the temperature profiles calculated with different model assumptions. 
Variations of $d_{pl}$ and $l_{pl}$ do not greatly affect the peaking of the radial, subsequently, axial temperature profiles. 
At the same time, replacing the Gaussian $Q\left(z\right)$-profile with the flat profile 
leads to flattening of the calculated axial temperature profile as well, bringing its shape closer 
to that in the experiment. This result is an indication that a relatively flat $Q\left(z\right)$ distribution 
could bee indeed a more realistic assumption than the Gaussian profile applied in the reference model. 

\Table{\label{sensitivity:Tmax:T} Maximum temperature $T_{max}$, in kK,
calculated on different assumptions for selected model cases}
\br
  model case & ref. & flat Q(z) & $d_{pl}\cdot$1.5 & $d_{pl}$/1.5 & $l_{pl}\cdot$1.5 & $l_{pl}$/1.5 \\
\mr
 & \multicolumn{6}{c}{ 0.2 bar } \\ 
\mr
 0.9 kW, 20 slm (SEI=0.64 eV) & 8.4  & 7.9 & 5.6 & 10.0 & 6.7 & 9.6 \\ 
\mr
 2.7 kW, 20 slm (SEI=1.9 eV) & 10.4  & 9.9 & 8.5 & 13.5 & 9.4 & 11.8 \\ 
\mr
 & \multicolumn{6}{c}{ 0.9 bar } \\ 
\mr
 0.9 kW, 40 slm (SEI=0.32 eV) & 4.4 & 4.3 & 3.5 & 6.8 & 3.9 & 5.7 \\ 
\mr
 0.9 kW, 10 slm (SEI=1.3 eV) & 7.4 & 7.1 & 6.7 & 8.1 & 6.8 & 8.1 \\ 
\mr
 1.8 kW, 10 slm (SEI=2.5 eV) & 7.5 & 7.2 & 6.8 & 8.5 & 7.0 & 8.3 \\ 
\mr
 1.8 kW, 5 slm (SEI=5.1 eV) & 7.8 & 7.3 & 7.1 & 9.1 & 7.2 & 9.6 \\  
\br
\endTable

As one can see in figures~\ref{09bar:Tmax:fig}-\ref{02bar:Tmax:fig} 
the assumption that the real $d_{pl}$ is larger than its optical appearance  
can bring the calculated $T_{max}$ closer to the experimental values. 
This assumption would be in line with the observation made in~\cite{Viegas2021}. 
However, the 0.9~bar case with the lowest SEI (flow rate 40~slm) where the temperature calculated 
with increased $d_{pl}$ is too low would not fit into this hypothesis. 
Another factor which can strongly affect the calculated $T_{max}$ is the ionization 
which is missing in the model. This factor could be, in particular, responsible for too high temperatures at the lowest pressure 0.2~bar. 
Because of that known limitation of the applied model and because from the beginning that was not 
the intention of the present work it was not tried to fit the calculated temperature exactly into 
the measurements by adjusting the heat source shape. 

Despite large variation of $T_{max}$ in all cases listed in~\tref{sensitivity:Tmax:T} 
the conversion rates were found to vary only within $\pm$0.3~\% at maximum, 
staying in most cases unchanged within 3 decimal digits. 
This result can be also seen in  
figures~\ref{09bar:X:fig}-\ref{02bar:X:fig},~\ref{09bar:eta:fig}-\ref{02bar:eta:fig} 
where the effect of increasing and decreasing
$d_{pl}$ by a factor 1.5 is displayed by dashed and dotted lines respectively.
 
No good theoretical explanation was found for this insensitivity, and so far 
it has been accepted solely as an outcome of numerical experiments. 
In order to verify this result and to see in how far the conversion rate 
stays so irresponsive an extra test has been done for two 0.9~bar cases. 
The results are presented in~\tref{sensitivity:X:dpl:T}. 
One can see that the calculated $\chi$ starts to change only when $d_{pl}$ gets comparable 
to the discharge tube diameter ($d$=26~mm). At the same time, further increase of  
$d_{pl}$ may lead to near saturation as the solution approaches a 1D plug flow, 
as demonstrated especially by the second case in~\tref{sensitivity:X:dpl:T}. 

\begin{table}
\caption{\label{sensitivity:X:dpl:T} Conversion rate $\chi$ and maximum temperature $T_{max}$ as functions of plasma diameter $d_{pl}$ 
         in two model cases}
\begin{indented}
\item[] \begin{tabular}{@{}lll|lll}
\br 
\multicolumn{3}{c|}{ 0.9 bar, 10 slm, 0.9 kW } & \multicolumn{3}{c}{ 0.9 bar, 10 slm, 1.8 kW  }  \\ 
 \mr 
$d_{pl}$, mm & $\chi$, \% & $T_{max}$, K & $d_{pl}$, mm & $\chi$, \% & $T_{max}$, K  \\ 
 \mr 
 3.6 & 10.0 & 7410 &  5.5 & 12.4 & 7510 \\ 
 3.6$\times$1.5 & 10.0 & 6710 &  5.5$\times$1.5 & 12.4 & 6840 \\ 
 3.6$\times$2 & 10.0 & 5580 &  5.5$\times$2 & 12.4 & 6100 \\ 
 3.6$\times$3 & 9.7 & 3750 &  5.5$\times$3 & 12.4 & 3810 \\ 
 3.6$\times$4 & 8.9 & 3330 &  5.5$\times$4 & 12.2 & 3390 \\ 
 3.6$\times$5 & 7.8 & 3070 &  5.5$\times$5 & 12.0 & 3190 \\ 
 3.6$\times$6 & 6.6 & 2900 &  5.5$\times$6 & 11.8 & 3070 \\ 
\br 
\end{tabular}
\end{indented}
\end{table}

This observed insensitivity of the calculated conversion rate is in general a positive result 
which offers the possibility of relatively accurate calculation of $\chi$ (and $\eta$) 
despite large uncertainty in the assumed heat source shape. 

\subsection{Numerical insights into the conversion process}

\label{insights:S}

Since the agreement between the calculations and the experiment for 0.9~bar is good enough the 
model can be used to gain some insights into the conversion process 
which are not directly accessible by measurements. 
In particular, insights into the spatial effects. 
In very general terms the process can be separated in space into two regions. 
Into the 'plasma region' where the gas is heated and into the 'effluent region' 
downstream where the heating stops and the gas mixture cools down. 

\begin{figure}
\center
\subfloat[$z=$0~mm]{
\includegraphics[width=7cm]{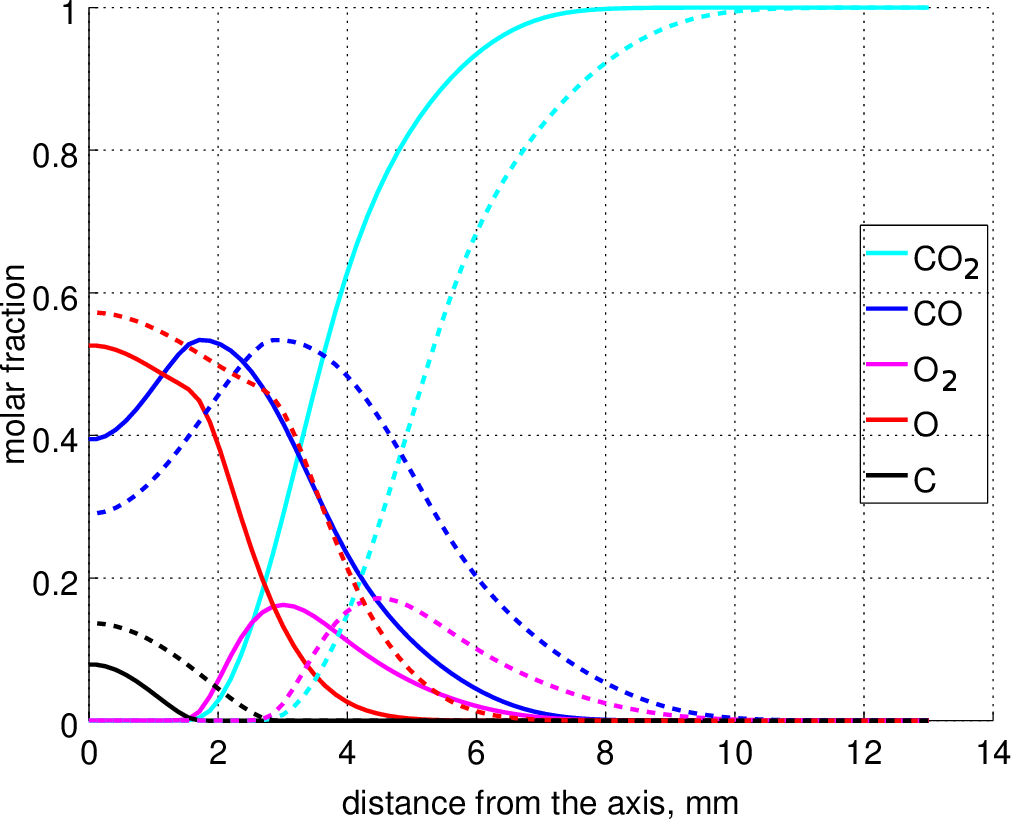} }
\subfloat[$z=$28~mm]{\label{molar:fractions:z2:fig} 
\includegraphics[width=7cm]{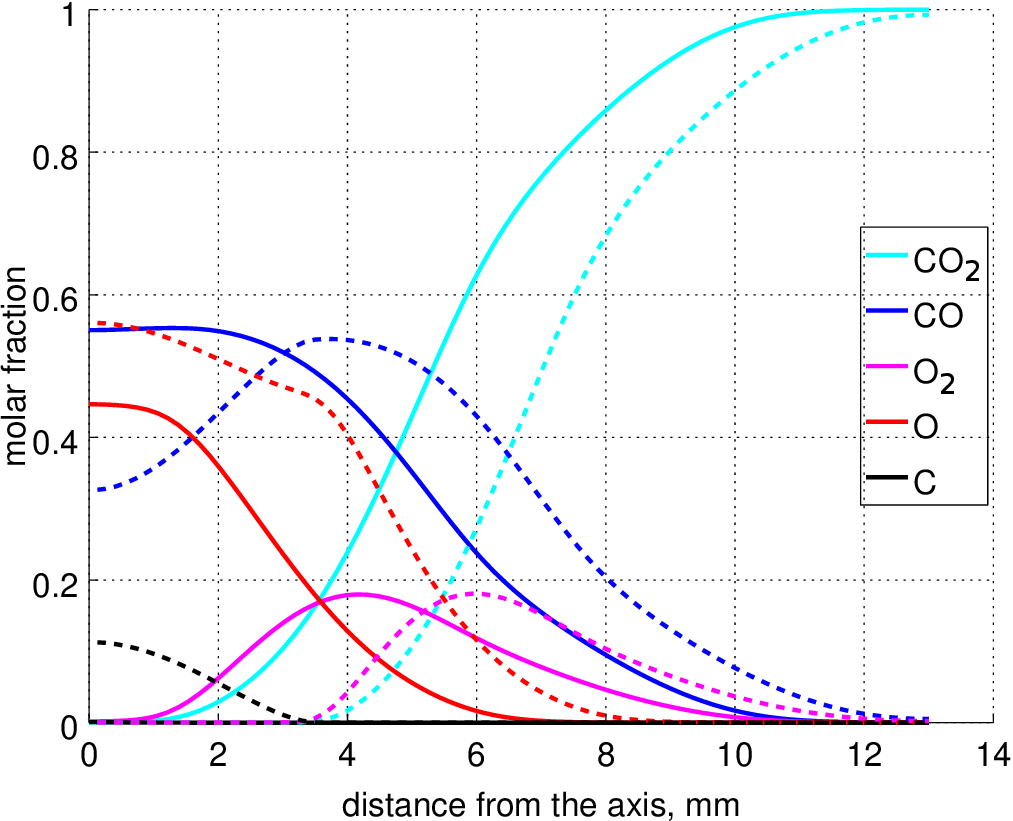} }
\caption{Radial profiles of the molar fractions: (a) is the middle of the heat source, 
         (b) is the same cross-section as in~\fref{temperature:rad:fig}. 
         Case 0.9~bar, 10~slm, solid lines: 0.9~kW, dashed lines: 1.8~kW}
\label{molar:fractions:fig} 
\end{figure}

The composition of the hot mixture is illustrated in~\fref{molar:fractions:fig} 
which shows the calculated radial profiles of the molar fractions in the middle of the heat source ($z$=0)
and near the  boundary between the 'plasma' and the 'effluent' ($z$=28~mm). 
This latter is the same cross section as that in~\fref{temperature:rad:fig}, 
and the nominal case shown by solid lines in~\fref{molar:fractions:fig} is exactly same as that in~\fref{temperature:fig}.
In addition, a case with twice as high input power is shown by dahsed lines.
Expectedly, the gas in the center is fully dissociated and consists only of CO/O/C. 
The profiles in the higher power case are broader because of larger assumed plasma diameter, 
$d_{pl}$=3.6~mm for 0.9~kW and $d_{pl}$=5.5~mm for 1.8~kW.  
One can see that compared to the $z=0$ cross section at the end of the plasma region 
the molar fractions of CO$_2$ and O$_2$ in the center and of CO and O at the edge are increased. 
This increase is to large extent the result of radial diffusion. 
Significance of the diffusion process was revealed quantitatively by examining the particle 
balance in each individual flux tube. This examination showed that for all chemical components the 
contribution of the effective source due to the radial part of ${\rm div}{\left(\vec{\Gamma}_i\right)}$, 
see~\eref{diffusion:Eq}, is always comparable to that of the volumetric source $S_i$.

\begin{figure}
\center
\subfloat[0.32 eV (40 slm, 0.9 kW)]{\label{cum:eta:sei=03:fig} 
\includegraphics[width=5cm]{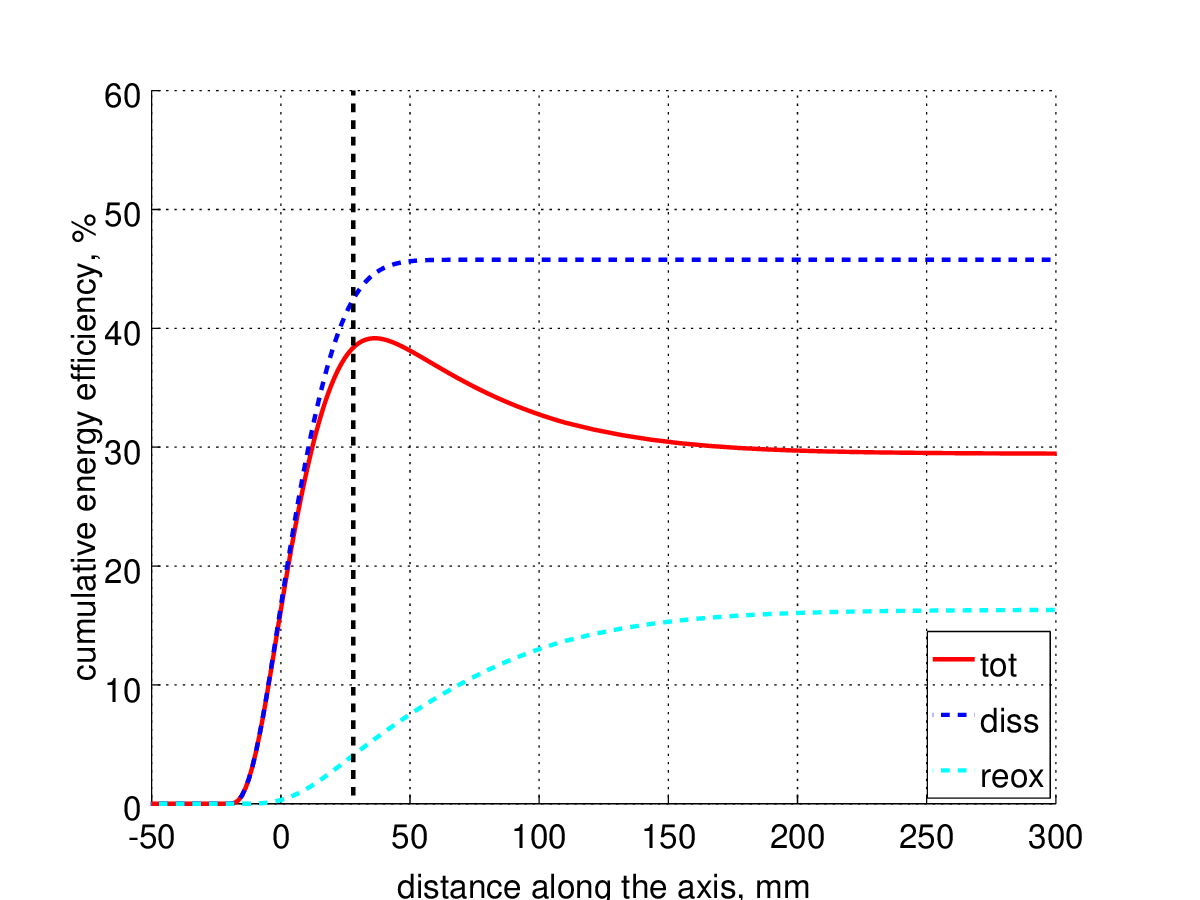} }
\subfloat[1.3 eV (10 slm, 0.9 kW)]{\label{cum:eta:sei=13:fig} 
\includegraphics[width=5cm]{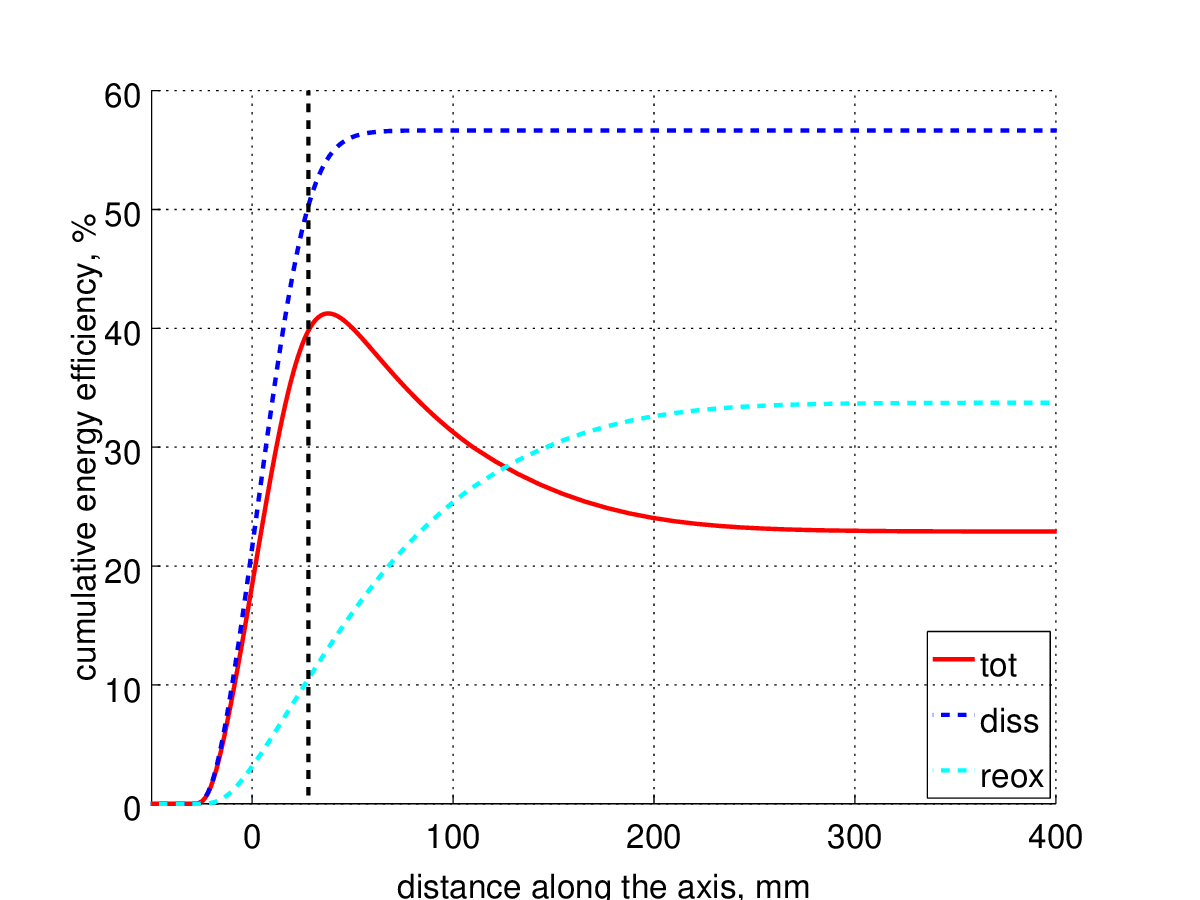} } 
\subfloat[2.5 eV (10 slm, 1.8 kW)]{\label{cum:eta:sei=25:fig} 
\includegraphics[width=5cm]{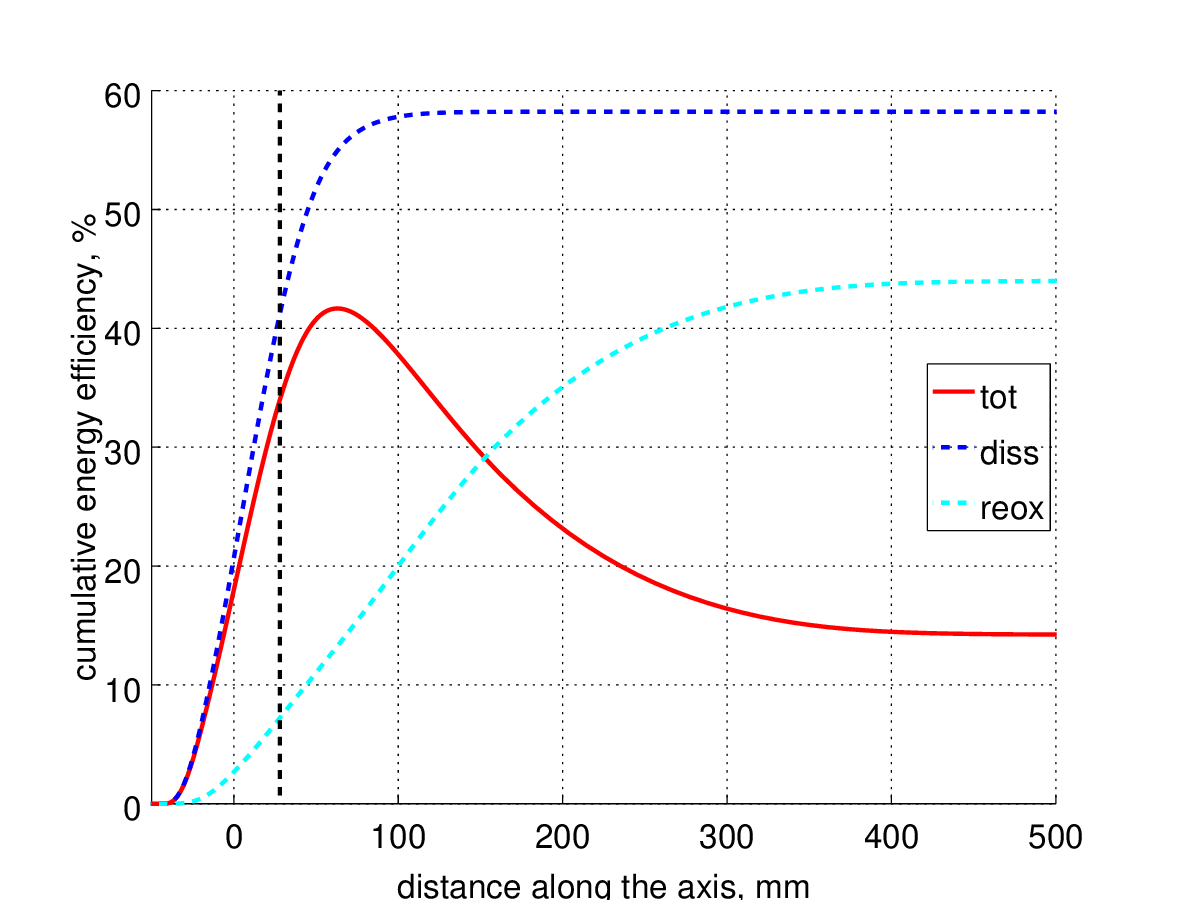} }
\caption{Examples of the cumulative local energy efficiency for different values of SEI at pressure 0.9~bar. 
         'tot' is $\eta\left(z\right)$, equation~\eref{SEI:definition:eq}, 
         'diss' and 'reox' are $\eta_{diss}\left(z\right)$ and 
         $\eta_{reox}\left(z\right)$,~\eref{eta:diss:eq} and~\eref{eta:reox:eq} respectively. 
         Vertical dashed line is the cross-section $z=$28~mm, same as in~\fref{temperature:rad:fig} 
         and~\ref{molar:fractions:z2:fig} }
\label{cum:eta:fig} 
\end{figure}

The CO produced in the plasma region partly recombines back into CO$_2$ in 
the effluent region due to the reverse reactions of the processes N1, N2 (\tref{chemical:mechanism:T}).
This backward transformation determines the final net effect of the conversion process. 
Its impact on the overall energy efficiency 
can be illustrated by introducing the 'cumulative local efficiency' $\eta\left(z\right)$ 
defined by modifying~\eref{conversion:definition:eq},~\eref{efficiency:definition:eq} as follows:
\begin{equation}
 \eta\left(z\right) = 
 \frac{\Delta H_f }{\mbox{SEI} }\cdot 
 \left( 1 - \frac{  2\pi\int^{d/2}_0 \left[ \Gamma_{CO_2}\left(z,r\right)\right]_z r dr }{ \mbox{flow rate} } \right)
 \label{cumulative:eta:eq}
\end{equation}
Examples of the $\eta\left(z\right)$ profiles for selected cases with different SEI are shown in~\fref{cum:eta:fig}.
The maximum of $\eta\left(z\right)$ is reached near the end of the plasma region, 
and its absolute value is almost same in all three cases - around 40~\%.  
Note that the assumed plasma length in the last case $l_{pl}$=81~mm is larger than in the two previous cases 
($l_{pl}$=55~mm), therefore the position of the maximum is shifted to the right. 
The reduction of $\eta\left(z\right)$ after reaching the peak value gets stronger with increased SEI 
which produces the net effect seen in~\fref{09bar:eta:fig}. 

Increase of the total axial CO$_2$ flux between the end of the plasma region and the outlet obtained in 
the model agrees qualitatively with the experimental results reported in~\cite{Wiegers2022}.
In~\cite{Wiegers2022} a microwave discharge similar to that considered in the present paper was investigated, 
and the conversion rates were measured at two locations: near the plasma (downstream) 
and at the outlet. It was demonstrated that the conversion rates determined near the 
plasma region are indeed much larger than the final rates at the outlet. 

\begin{figure}
\center
\subfloat[dissociation zone]{
\includegraphics[width=8cm]{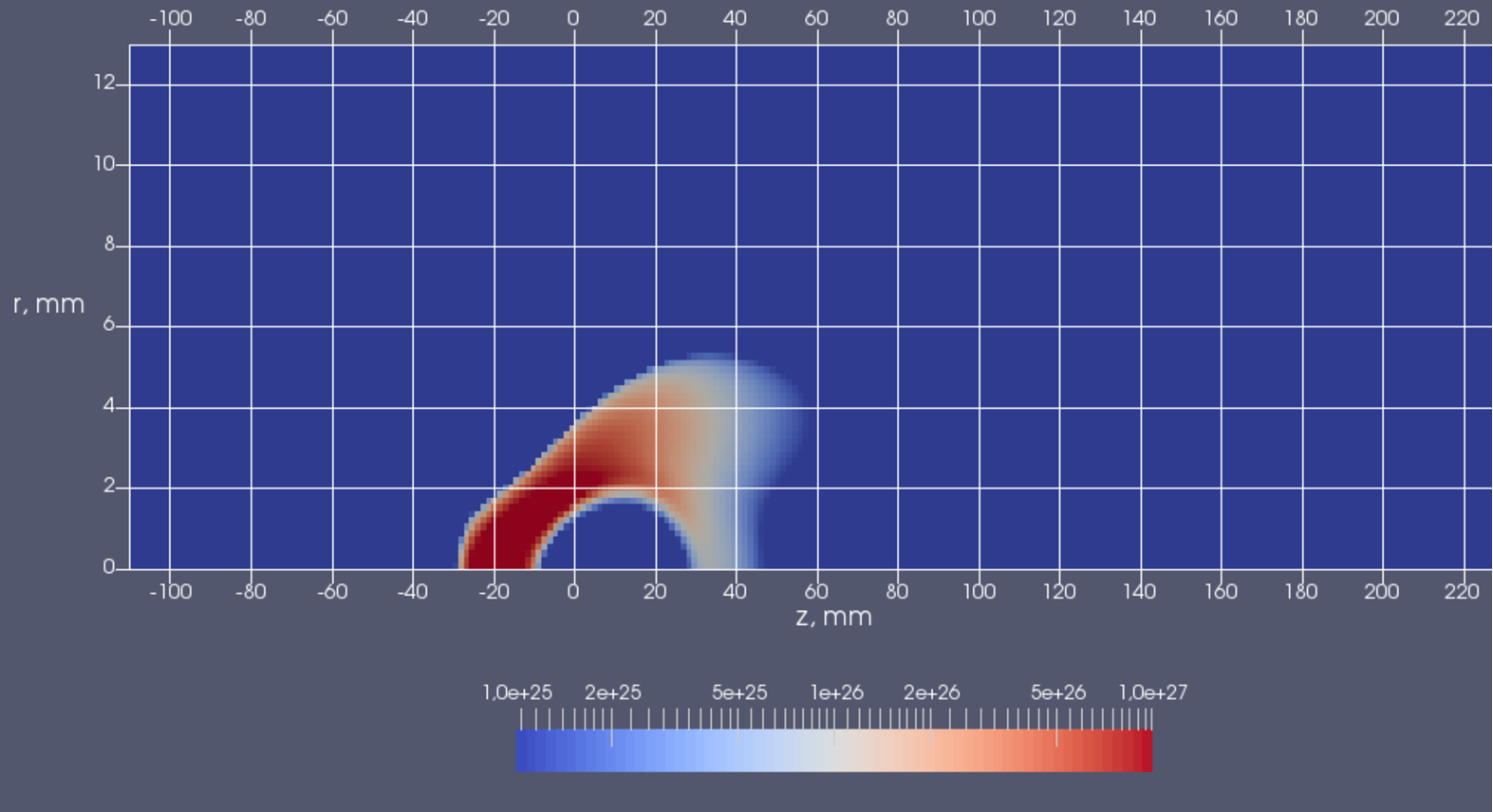}}\\
\subfloat[re-oxidation zone]{\label{reox:fig}
\includegraphics[width=8cm]{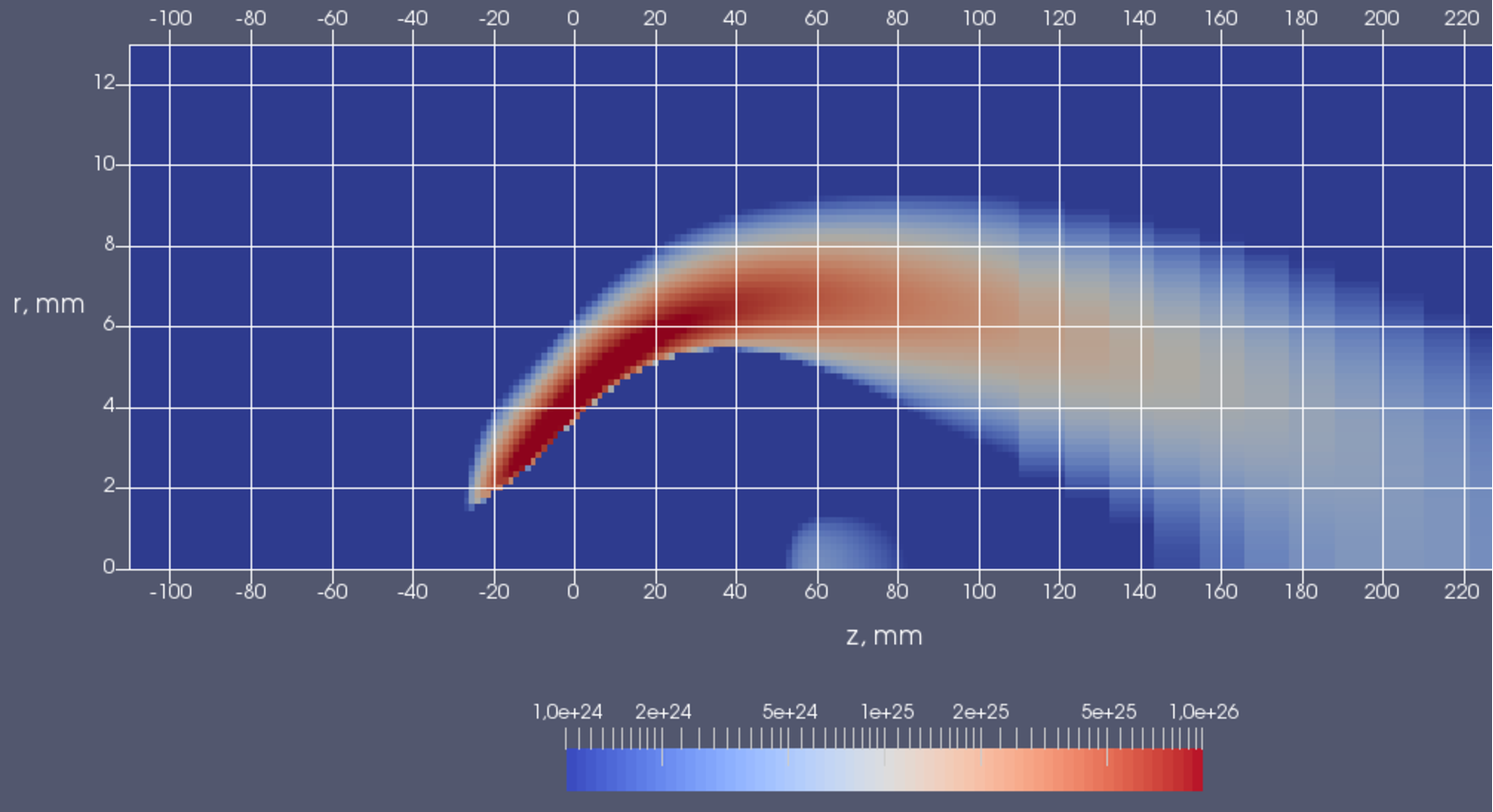}}
\caption{Spatial distribution of the (a) net CO$_2$ sink, $\max{\left[-S_{CO_2}\left(z,r\right),0\right]}$, m$^{-3}$/s;  
         (b) net CO$_2$ source, $\max{\left[S_{CO_2}\left(z,r\right),0\right]}$, m$^{-3}$/s. Case 0.9~bar, 10~slm, 0.9~kW}
\label{diss:reox:fig}         
\end{figure}

The one-dimensional representation of~\fref{cum:eta:fig} can be further refined by considering two zones. 
The 'dissociation zone' where the net local volumetric source of CO$_2$ is negative, $S_{CO_2}\left(z,r\right)<$0, 
and the 're-oxidation' zone where  $S_{CO_2}\left(z,r\right)>$0
\footnote{Here the term 're-oxidation' is preferred to 'recombination' because this latter 
         can be confused with recombination of free electrons in plasma}. 
An example of the spatial distribution of $S_{CO_2}$ in both zones is shown in~\fref{diss:reox:fig}.
One can see that the re-oxidation zone is not concentrated downstream of the plasma region, but is rather 
spread in the radial direction. This example is another illustration of the essential 
two-dimensionality of the conversion process -  in line with the 'cold core / hot shell' consideration 
put forward in~\cite{denHarder2017} and further elaborated in~\cite{Wolf2020a}.

Individual contributions of the two zones can be added to the diagrams of~\fref{cum:eta:fig} by introducing the following factors:
\begin{equation}
 \eta_{diss}\left(z\right) =  \frac{\Delta H_f }{\mbox{SEI} }\cdot 
 \frac{ 2\pi\int^z_{{\rm inlet}}\int^{d/2}_0 \max{\left[-S_{CO_2}\left(z',r\right),0\right]}rdrdz'}{ \mbox{flow rate} } 
 \label{eta:diss:eq}
\end{equation}
\begin{equation}
 \eta_{reox}\left(z\right) =  \frac{\Delta H_f }{\mbox{SEI} }\cdot 
 \frac{ 2\pi\int^z_{{\rm inlet}}\int^{d/2}_0 \max{\left[S_{CO_2}\left(z',r\right),0\right]}rdrdz'}{ \mbox{flow rate} } 
 \label{eta:reox:eq}
\end{equation}
It is readily seen that due to particle conservation $\eta\left(z\right) = \eta_{diss}\left(z\right) - \eta_{reox}\left(z\right)$. 

\Fref{reox:fig} suggests that substantial re-oxidation can take place already at the edge of the plasma region, 
and not only in the effluent. 
The profiles of $\eta_{reox}\left(z\right)$ plotted as dashed lines in~\fref{cum:eta:fig}
confirm that the CO losses there lead to substantial reduction of the peak $\eta\left(z\right)$.
This latter, if determined solely by $\eta_{diss}\left(z\right)$, could have been reached almost 60~\% in the maximum. 
Nevertheless, one can also 
clearly see that the major negative impact of the re-oxidation which is responsible
for the drop of $\eta$ with increased SEI is localized downstream from the plasma region in the effluent. 

\section{Conclusions}

\label{conclusions:S}

The main idea of the thermo-chemical approach for modelling of the CO$_2$ conversion in 
microwave flow reactors is that the plasma only enters the model as prescribed fixed heat source 
whose spatial localization is taken from experiment. 
The mathematical problem is then reduced to modelling of a flow of hot gas with chemical reactions. 
In the present paper this approach has been benchmarked against experimental data of 
F.~A.~D’Isa et al.~\cite{DIsa2020,DIsaThesis2021} by comparing them with 1.5D numerical transport calculations.
The 1.5D transport model solves 2D equations for the heat balance and 
for particle balance of each chemical component 
and uses local 1D approximation for the axial flow. 
A basic physico-chemical model of the CO$_2$/CO/O$_2$/O/C mixture has been proposed 
which comprises the minimal reaction mechanism and molecular transport coefficients 
without turbulent transport. The comparison has been performed for pressures 0.9, 0.5 and 0.2~bar. 
The specific energy input per molecule SEI for 0.9~bar was varied between 0.3 and 5~eV, 
and for two other pressures between 0.6 and 2~eV.

The calculated conversion rates (and energy efficiencies) are found to be always in good agreement with 
experiment, mostly within experimental error bars. At the same time, deviations between the calculated 
and measured gas temperatures are always large. For 0.9~bar and 0.5~bar the maximum temperature $T_{max}$ 
is in most cases overestimated by the reference model by 20-30~\%. 
Comparison of the radial and axial temperature profiles show that besides the deviation of 
absolute values the calculated profiles appear more peaked than the experimental ones. 
For 0.2~bar the calculated $T_{max}$ can exceed the experimental value by almost a factor 2 getting 
significantly larger than 8000~K which makes the results nonphysical. Nevertheless, even in those cases the calculated 
conversion rates agree well with the measurements.

The thermo-chemical model requires the characteristic diameter and length
of the heating zone $d_{pl}$ and $l_{pl}$ as input parameters. 
They are estimated in experiment from the intensity of plasma radiation 
and have relatively large uncertainty. 
Therefore, sensitivity of the numerical solutions with respect to increasing and reducing 
$d_{pl}$, $l_{pl}$ by a factor 1.5 has been investigated. 
It has been found that the calculated $T_{max}$ and the temperature profiles 
are sensitive with respect to the assumed $d_{pl}$ and $l_{pl}$, but the conversion rates are not sensitive. 
Extra tests with further increased $d_{pl}$ have demonstrated that the calculated conversion rate 
stats to change only when $d_{pl}$ gets comparable to the diameter of the discharge tube. 

The results of the present work confirm that the conversion of CO$_2$ 
observed in contracted microwave plasmas can be explained by a purely thermo-chemical mechanism  
which coincides with the analysis presented previously in~\cite{Bongers2017,denHarder2017,DIsa2020,vandenBekerom2019,Wolf2020a}.
The model backs the concept that the process is essentially two-dimensional 
and the energy efficiency is mainly limited by backward reactions ('re-oxidation') in the effluent downstream from the 
plasma region. 

The 1.5D model applied here can be used for fast evaluation of the physico-chemical models 
in experiments with flow in straight channels. 
This transport model can be easily implemented numericaly by applying, e.g. the well known method of lines 
(see~e.g.~\cite{ReactingFlowBook2003},~Chapter~7.5). 
The 1.5D model is not applicable to the flows through channels of complicated geometry with 
nozzles where the basic underlying assumptions, in particular the assumption of constant pressure,
are most likely not fulfilled anymore. 
For such flows more complete numerical tools which solve the full Navier-Stokes equation are required, 
but the  physico-chemical model described in section~\ref{phys:chem:model:S} is applicable in those cases too. 
The present results suggest that this model can be used for practical calculations for pressures $\geq$0.5~bar.

\section*{Data availability statement}

The data that support the findings of this study are available from the corresponding author upon reasonable request. 
The source code of the physico-chemical model of the CO$_2$/CO/O$_2$/O/C mixture is available under the link:
\url{https://jugit.fz-juelich.de/thermochemistry/physchem-co2-co-o2-o-c}

\ack

Specific attributions of author contribution and responsibility. 
V~Kotov: Conceptualization, Formal Analysis, Methodology, Software, Validation, Visualization, Writing - original draft. 
Ch~Kiefer: Investigation, Writing - review \& editing. 
A~Hecimovic: Data curation, Investigation, Writing - review \& editing
\appendix

\section{Numerical implementation}

\label{numerics:A}

Numerical solution of~\eref{diffusion:Eq},~\eref{heat:Eq},~\eref{flow:Eq} is implemented in a self-written Fortran 2003 code
which applies finite-volume method on regular orthogonal grids~\cite{Patankar1980}. 
The stationary solution is found by adding time-derivatives and iterating the time-dependent 
problem until convergence. Fully implicit scheme is applied on each time-iteration. 
The solver SSLUGM from the SLAP package~\cite{SLAP} was used to solve the sets of linear equations
with sparse matrix (preconditioned Krylov subspace methods based on the generalized minimum residual method).

Patankar's implicit under-relaxation~\cite{Patankar1980}, Section 4.5 there, 
and source term linearization~\cite{Patankar1980}, Section~7.2-2, are applied to increase numerical stability. 
Equations~\eref{diffusion:Eq} are solved only for $N_s-1$ components, where $N_s$ is the total number of 
components. The molar fraction of one component which is chosen to be the 'main' one $X_{i'}$ is found 
as $X_{i'} = 1 - \sum_{i\ne i'} X_i$. CO$_2$ seems to be the natural choice for $i'$. 
However, experience has shown that choosing CO as the main component greatly improves numerical stability and accuracy. 

The whole phsyco-chemical model of~\sref{phys:chem:model:S} is implemented completely independent from the solver
as one Fortran module with separate unit-tests.
The classes which implement spatial discretization were unit-tested by the Method of Manufactured Solutions. 
Heat transfer problems with the analytic solutions one-dimensional in $z$ or $r$ direction were used 
for integrated tests of the solver of the whole system~\eref{diffusion:Eq},~\eref{heat:Eq},~\eref{flow:Eq}.  
With one single component as well as with several identical components (which must produce exactly same solution). 
The convergence is monitored by controlling residuals of~\eref{diffusion:Eq},~\eref{heat:Eq} 
and global balances. In all calculations discussed in the paper the errors in the 
global particle balance of any chemical component are always smaller than $2\cdot10^{-3}$ of the 
total flow rate, and the errors in the global energy balance are always smaller than $10^{-3}$ 
of the total input power. The calculations were performed on relatively coarse grids with 120$\times$40 cells. 
Grid size independence of the solution was checked by repeating the 6 cases listed in~\tref{sensitivity:Tmax:T} 
on larger 240$\times$80 cells grids.

Flows with chemical reactions are mathematically stiff problems. To simulate them usually special operator splitting 
schemes are applied~\cite{Boesenhofer2023}. This was not done in the present work - with negative 
consequences for numerical performance. To reach the level of convergence of the global balances reported above 
the time-step 10$^{-5}$~sec or smaller was required. At the same time, to achieve the steady-state 
for the problems in question a simulation had to run over 0.1-10~sec of physical time. 
As a result, one simulation even on 120$\times$40 grid takes several hours or more. 
This experience emphasizes that applying appropriate operator-splitting, 
such as the popular Wu-scheme~\cite{Wu2019}, is highly desirable for practical simulations 
of thermo-chemical plasma conversion even if only a relatively small number of components - 5 in the present case - 
are taken into account. 

\section{Extract from kinetic theory}

\label{kin:theory:A}

According to~\cite{Hirschfelder1954}, Equations~(8.2-9),~(8.2-11), 
the following expressions can be derived for the coefficient 
of self-diffusion $\mathcal{D}$ and thermal conductivity $\lambda$ 
of a single component gas of spherically-symmetric molecules without internal degrees of freedom:
\begin{equation}
  \mathcal{D} = \frac{3}{8} \frac{\sqrt{\pi mk T}}{\pi m n \sigma^2 \Omega^{(1,1)*}}
  \label{kin:theory:D:eq}
\end{equation}
\begin{equation}
 \lambda = \frac{25}{32} \frac{ \sqrt{\pi m k T} }{ \pi\sigma^2 \Omega^{(2,2)*} } \frac{c_v}{m}
 \label{kin:theory:lam:eq}
\end{equation}
In those expressions $\sigma$ is the effective rigid sphere collision cross-section, 
$c_v$ is the molar heat capacity at constant volume,  $\Omega^{(1,1)*}$ and 
$\Omega^{(2,2)*}$ are the normalized collision integrals. Those latter  
are the total collision integrals $\Omega^{(1,1)}$, $\Omega^{(2,2)}$
defined in~\cite{Hirschfelder1954}, \S~7.4d, Equation (7.4-34), divided by the 
collision integrals calculated for the rigid spheres type of interaction for which the integration can 
be performed analytically: \cite{Hirschfelder1954}, Equation (8.2-8).  

Combining~\eref{kin:theory:D:eq} and~\eref{kin:theory:lam:eq} yields the relation between $\lambda$ and $\mathcal{D}$:
\begin{equation}
\lambda = \frac{25}{12} \frac{c_v}{A^*}n\mathcal{D}
\label{mu:lam:D:eq}
\end{equation}
where $A^* = \frac{ \Omega^{(2,2)*}}{\Omega^{(1,1)*}}$ is shown to be nearly independent of $T$.
Results of the calculation of $A^*$ for Lennard-Jones potential can be found in~\cite{Hirschfelder1954}, Table~I-N. 
They demonstrate that in a very wide temperature interval $A^*$ only varies between 1.05 and 1.15. 
Therefore, in the calculations always $A^*$=1.1 is taken. The accuracy provided by~\eref{mu:lam:D:eq} 
with $\mathcal{D}$ calculated by Fuller scaling (\cite{PropertiesGasesLiquids5th},~Equation~(11-4) there) 
was tested by applying it to calculate $\lambda$ for O-atoms for which 
the accurate quantum-mechanical calculations~\cite{Holland1988} are available. 
This test has shown that in the temperature range 2000..10000~K this method 
overestimates the results of~\cite{Holland1988} by 40~\% at maximum, which is a reasonable agreement.

\section*{References}

\end{document}